\documentclass[12pt]{article}

\pdfoutput=1

\input epsf
\usepackage{amssymb,amsmath,wrapfig,subfigure,pifont}
\usepackage[matrix,arrow,curve]{xy}
\usepackage{cite}
\usepackage{graphicx,color}
\usepackage[debug,pageanchor=false]{hyperref}
\definecolor{link}{rgb}{.8,.15,.1}
\hypersetup{colorlinks=true,linkcolor=link,citecolor=link,urlcolor=link,linktocpage}
\usepackage{cancel}
\usepackage{comment}

\makeatletter
\@addtoreset{equation}{section}
\makeatother

\def\a7 {\mathrm{AdS}_7}
\def\a4 {\mathrm{AdS}_4}

\def\nn {{\cal N}}

\def\rr {{\mathbb{R}}}
\def\zz {{\mathbb{Z}}}
\def\del {\partial}

\def\del {\partial}

\def\stt {{$\mathrm{SU(3)}\times\mathrm{SU(3)}$}}

\def\lsim{\mathrel{\rlap{\lower4pt\hbox{\hskip1pt$\sim$}}
    \raise1pt\hbox{$<$}}}                

\def\Re{\mathrm{Re}}
\def\Im{\mathrm{Im}}

\begin{document}

\begin{titlepage}

\begin{center}

\vskip .3in \noindent

{\Large \bf{AdS$_4$ compactifications of \\\vspace{.1cm} AdS$_7$ solutions in type II supergravity}}

\bigskip

	Andrea Rota and Alessandro Tomasiello\\

	\bigskip
	  Dipartimento di Fisica, Universit\`a di Milano--Bicocca, I-20126 Milano, Italy\\
	and\\
	INFN, sezione di Milano--Bicocca,
	I-20126 Milano, Italy

	\vskip .3cm
	{\small \tt andrea.rota@unimib.it, alessandro.tomasiello@unimib.it}
	\vskip 2cm
	     	{\bf Abstract }
	\vskip .1in
\end{center}

We find new classes of AdS$_4$ solutions with localized branes and orientifolds, both analytic and numerical. We start with an Ansatz for the pure spinors inspired by a recently found class of AdS$_7\times M_3$ solutions in massive IIA; we replace the AdS$_7$ by AdS$_4\times \Sigma_3$, and we fibre $M_3$ over $\Sigma_3$ in a way inspired by a field theory SU(2) twist. We are able to reduce the problem to a system of five ODEs; a further Ansatz reduces them to three. Their solutions can be bijectively mapped to the AdS$_7$ solutions via a simple universal map. This also allows to find a simple analytic form for these solutions. They are naturally interpreted as twisted compactifications of the $(1,0)$ CFT$_6$'s dual to the AdS$_7$ solutions. The larger system of five ODEs also admits more general numerical solutions, again with localized branes; regularity is achieved via an attractor mechanism.

\noindent

\vfill
\eject

\end{titlepage}

\hypersetup{pageanchor=true}

\tableofcontents

\section{Introduction}

In many string theoretic constructions, the presence of extended sources such as D-branes or O-planes is a crucial ingredient. In compactifications, for examples, O-planes are thought to be important to overcome no-go arguments that forbid de Sitter (or even Minkowski with non-trivial flux) compactifications \cite{gibbons-nogo,haridass-dewit-smit,maldacena-nunez}. However, in most cases these sources back-react on the metric in a way which destroys whatever symmetries were previously present, and makes it prohibitively hard to find a full solution to the equations of motion. 

To overcome this problem, sources are often ``smeared'' over the internal space: namely, they are assumed to occur in a continuous distribution with varying positions, much like the individual electrons on a charged piece of conductor. While this is fine for D-branes, it is incompatible with the definition of an O-plane, which must in fact lie at the fixed locus of an involution. When the smearing trick is performed on O-planes, it is usually done with the hope that it might be a good indicator of whether a non-smeared solution exists. It is hence interesting to find solutions with localized (i.e.~non smeared) sources, even ones where the cosmological constant is negative. Although there already exists one family of supersymmetric AdS$_4$ solutions with localized sources, in type IIB supergravity \cite{assel-bachas-estes-gomis}, such examples remain rare.

In this paper, we are going to present a class of infinitely many new supersymmetric AdS$_4$ solutions with localized sources, in type IIA supergravity with Romans mass parameter $F_0$. As an example: 
\begin{equation} \label{eq:intro-D6}
	ds^2_{10} = \frac {5^{3/2}}{12\sqrt{2}}\frac{n_2}{F_0}\sqrt{\tilde y + 2} \left[ ds^2_{\rm AdS_4} + \frac{4}{5} ds^2_{\Sigma_3} 
	+ \frac{3}{10} \frac{d \tilde y^2}{(1-\tilde y)(\tilde y+2)} + \frac{4}{5} \frac{(1-\tilde y)(\tilde y+2)}{\tilde y^2-5 \tilde y+10} Ds^2_{S^2} \right] ,
\end{equation}
with $\tilde y \in [-2,1]$, $\Sigma_3$ a compact hyperbolic three-manifold, and $Ds^2_{S^2}$ the round $S^2$ metric fibred over $\Sigma_3$ in a certain way. This has a stack of $n_2$ D6-branes at $\tilde y=-2$, and it is regular at $\tilde y=1$, so that the topology of the space $M_3$ described by $\tilde y$ and the $S^2$ is that of an $S^3$. We will also present analytic solutions with two D6 stacks, with O6 singularities, and with D8-branes. Moreover, we will present numerical solutions where $\Sigma_3$ can be replaced with an $S^3$, and also where sources can even be absent; in particular we will have a family of completely regular solutions with topology AdS$_4\times S^3\times S^3$, but different from the one in \cite{behrndt-cvetic}.

Let us now explain how these solutions came about. Recently, a class of supersymmetric AdS$_7$ solutions was found \cite{afrt} where several types of localized sources were present. In that paper and in the follow-up \cite{gaiotto-t-6d}, more attention was given to solutions with only D8-branes (actually, D8/D6 bound-states), with an eye to the study of their holographic duals. However, solutions with localized D6's also exist; in \cite[Sec.~5.2]{afrt} one example was given, where one stack of D6's was possible. As was hinted there and we will see more explicitly here, it is also possible to have two stacks (with unequal numbers of D6's), or also to have O6's. Perhaps most striking was the fact that such localized sources were not hard to find: the system of ODEs got attracted to either D6 or O6 type of singularities, and it was in fact their absence that required fine-tuning. (Imposing that the number of D6's is integer did however require fine-tuning.)

In view of the issues explained above with localized branes, it was then interesting to ask whether those findings could be somehow transported to four dimensions. (Indeed, in a series of interesting papers \cite{blaback-danielsson-junghans-vanriet-wrase-zagermann,gautason-junghans-zagermann,blaback-danielsson-junghans-vanriet-wrase-zagermann-2,junghans-schmidt-zagermann}, an AdS$_7\times M_3$ setup similar to \cite{afrt} was examined to understand the differences between localized and smeared branes.) For this, we needed to somehow replace AdS$_7$ with AdS$_4\times \Sigma_3$, where $\Sigma_3$ is some new compact three-manifold. 

From a holographic point of view, this sounds like compactifying the CFT$_6$ to a CFT$_3$, on a three-manifold $\Sigma_3$. This is more commonly done from a CFT$_6$ to a CFT$_4$, thus replacing AdS$_7$ with AdS$_5\times \Sigma_2$. A famous example is the Maldacena-Nu\~nez solution \cite{maldacena-nunez}, which is dual to a ``twisted'' compactification of the $(2,0)$ theory on a Riemann surface. But it is indeed also possible to compactify on a three-manifold: the solution dual to this is in fact even older, going back to \cite{pernici-sezgin} (later being lifted to eleven dimensions in \cite{acharya-gauntlett-kim,gauntlett-macconamnha-mateos-waldram}).

Inspired by \cite{pernici-sezgin,gauntlett-macconamnha-mateos-waldram}, we formulated an Ansatz which would be holographically dual to compactifying the $(1,0)$ CFT$_6$ on a compact quotient of a  maximally symmetric $\Sigma_3$. We then used this Ansatz in the generalized complex geometry formalism of \cite{gmpt2}, where AdS$_4\times M_6$ solutions of type II supergravity were reformulated in terms of certain ``pure spinor equations''. With the Ansatz we formulated, these equations reduce to five ODEs, for five functions (the dilaton and warping, and three coefficients of the metric) depending basically on the coordinate $\tilde y$ in (\ref{eq:intro-D6}). 

We found two classes of solutions to the five ODEs, which we call respectively ``natural'' and ``attractor'' solutions. The ``natural'' class comes about when we notice that a certain three-dimensional subspace of the parameter space is left invariant by the flow defined by the ODEs. In other words, with a certain constraint the ODEs reduce to three; this requires assuming that $\Sigma_3$ be hyperbolic,\footnote{Compactifying the $(1,0)$ theories of \cite{hanany-zaffaroni-6d,brunner-karch} on a torus $T^3$ should also be possible, but presumably this leads to a solution that looks singular in IIA, and whose more appropriate description is in type IIB; it should correspond to the solutions in \cite{assel-bachas-estes-gomis}, which are dual to CFT$_3$'s obtained from Hanany--Witten configurations \cite{hanany-witten}.} but it simplifies the problem quite a bit. In fact, at this point we recognize that the three ODEs were quite similar to the ones given in \cite[Eq.(4.16)]{afrt} for AdS$_7$ solutions. This allows us to find a one-to-one correspondence between our natural class of AdS$_4$ compactifications and the AdS$_7$ solutions of \cite{afrt,gaiotto-t-6d}. At the level of the metric, the map reads
\begin{equation}\label{eq:intromap}
\begin{split}
	e^{2A}ds^2_{\rm AdS_7}+ &dr^2 + e^{2A}v^2 ds^2_{S^2}\ \to \ \\
	&\left( \frac{5}{8} \right)^{3/2} \left[ e^{2A} \left( ds^2_{\rm AdS_4} + \frac{4}{5} ds^2_{\Sigma_3} \right) + \frac85 \left( dr^2 + \frac{v^2}{1-6v^2} Ds^2_{S^2}\right) \right] \ ,	
\end{split} 
\end{equation}
where $A$, $v$ are functions of $r$, and $ds^2_{S^2}$ is the round metric on the $S^2$, which after the map gets fibred over $\Sigma_3$ in a certain way. There are infinitely many AdS$_7$ solutions, with arbitrary numbers of D8's, whose numbers and charges can be labeled by two Young diagrams \cite[Sec.~4]{gaiotto-t-6d}. So an immediate consequence of (\ref{eq:intromap}) is that we have an infinite number of AdS$_4$ solutions as well. As we mentioned already, there are also AdS$_7$ solutions with D6's and O6's, which were only quickly mentioned in \cite{afrt}; under the map (\ref{eq:intromap}), these become AdS$_4$ solutions which also have those sources. 

Moreover, a parallel paper \cite{afpt} studies compactifications where the AdS$_7$ in \cite{afrt} is replaced by AdS$_5\times \Sigma_2$, with $\Sigma_2$ a Riemann surface --- more similarly to the original Maldacena--Nu\~nez solution \cite{maldacena-nunez}. A version of (\ref{eq:intromap}) also holds for that case; see \cite[Eq.(1.3)]{afpt}. Crucially, in that paper the AdS$_5$ solutions were found analytically. This allows also to find an analytic expression to the AdS$_7$ solutions of \cite{afrt,gaiotto-t-6d}, and then using (\ref{eq:intromap}) to find analytic expressions for our AdS$_4$ case. This is how we found (\ref{eq:intro-D6}). 

The holographic duals of the AdS$_7$ solutions in \cite{afrt,gaiotto-t-6d} were argued in \cite{gaiotto-t-6d} to be CFT$_6$'s arising from NS5-D6-D8 brane configurations studied long ago \cite{hanany-zaffaroni-6d,brunner-karch}. By construction, our AdS$_4$ solutions will then be dual to the compactifications of those CFT$_6$'s on hyperbolic three-manifolds $\Sigma_3$. It would be interesting to understand what these CFT$_3$'s are; this might eventually lead to a generalization of the 3d-3d correspondence of \cite{dimofte-gaiotto-gukov}. (Notice however that supersymmetry is lower, namely ${\cal N}=1$.)

We mentioned that a certain constraint reduces the number of ODEs from five to three; this is what led us to the class of ``natural compactifications'' we talked about so far, the ones to which the map (\ref{eq:intromap}) applies. We are actually also able to make some numerical progress on the original system of five ODEs, obtaining another class which we call ``attractor solutions''. In this case, life is much harder: the system does not get attracted automatically to the physically sensible D6 and O6 singularities. Rather, if one evolves from the equator of $M_3\cong S^3$ towards the poles, in general one ends up with singularities which appear not to have any physical interpretation. However, with some inspiration from the natural case, we were able to identify boundary conditions which correspond to the presence of D6's and O6's: these consist in a certain perturbative solution in terms of fractional powers of the radial coordinate. These boundary conditions leave some free parameters, and it turns out that for an \emph{open set} in the space of these parameters the solution gets attracted in the other pole to a regular point. This works especially well for $\Sigma_3=S^3$, in a somewhat opposite fashion to the natural compactifications class. 

The paper is organized as follows. We start in section \ref{cft6sec} with a review about compactifications of CFT$_6$ and holography; this is background material in order to motivate our Ansatz in section \ref{sec:tech}. In that section we also review briefly the pure spinor techniques that we will use for supersymmetry. We will then analyze the solutions that we called natural in section \ref{natural}, and finally (in less detail) the ones we called attractor solution in section \ref{attractor}.

\section{CFT$_6$ compactifications in supergravity} \label{cft6sec}

As discussed in the introduction, in this paper we are interested in compactifying the $(1,0)$ CFT$_6$ of \cite{hanany-zaffaroni-6d,brunner-karch}, whose AdS$_7$ duals were found in \cite{afrt}. In order to formulate the correct Ansatz to achieve this, we will first review the compactifications of the $(2,0)$ theory and of its dual solution AdS$_7 \times S^4$. This is a widely explored subject that has led to great improvements in our understanding of the physics of M5 branes. 


If one puts a supersymmetric theory on a curved space without modifying its Lagrangian, supersymmetry will usually be broken by the curvature terms. Thus one needs to be careful about how one defines the theory on a curved space. An old strategy consists in a partial ``twist'' of the theory. Roughly speaking, fields with an R-symmetry index are considered to be sections of a certain R-symmetry bundle $E$, which is then chosen such that $E\otimes S$ (with $S$ the spinor bundle) admits a global section. This global section (which can then taken to be constant, up to a gauge transformation) is then a preserved supercharge. For brane theories, often the procedure also has a geometrical interpretation: $E$ can be interpreted as the normal bundle to the brane \cite{bershadsky-sadov-vafa}. Thus the twisting corresponds roughly to how one wraps the brane.

Compactifications of the M5 theory on Riemann surfaces $\Sigma_2$ were studied in \cite{maldacena-nunez} and more recently for example in \cite{gaiotto-maldacena,bah-beem-bobev-wecht}, both on the gravity and on the field theory side. There exist two possible ways of wrapping the M5s (i.e.~two different normal bundle geometries), which preserve eight or sixteeen supercharges. 

Compactifications on hyperbolic three-manifolds $\Sigma_3$ were studied in \cite{gauntlett-macconamnha-mateos-waldram}, lifting an earlier solution in \cite{pernici-sezgin}, preserving either four or eight supercharges.\footnote{Punctures along $\Sigma_3$ can also be introduced; they were studied in the probe approximation in \cite{bah-gabella-halmagyi}.} We will review these compactifications in section \ref{sub:11d}, and then rewrite them in terms of IIA supergravity in section \ref{sub:IIA}, with an eye to their generalization in presence of Romans mass.

\subsection{Compactifications from eleven-dimensional supergravity}\label{sub:11d}

In this section we will review $(2,0)$ compactifications in 11d supergravity, introducing notation that will be useful later when we will discuss the similar compactifications for the $(1,0)$ case. We will discuss the solutions only at the level of the metric. The spinorial supersymmetry parameters will be discussed in section \ref{sub:twsp}. 

The $(2,0)$ theory on the M5 worldvolume is dual to the AdS$_7 \times S^4$ background:
\begin{equation} \label{ads7s4}
ds^2_{11} = R^2 \left( ds^2_{\rm AdS_7} + \frac{1}{4} ds^2_{S^4} \right) \ . 
\end{equation}
Two types of compactifications on three-manifolds of this fully BPS background have been considered in the literature, preserving $\nn=1$ and $\nn=2$ supersymmetry in four dimensions. The $\nn=1$ solution corresponds to breaking of the SO(5) isometry group of the $S^4$ to SO(4), while in the $\nn=2$ case the subgroup preserved is SO(3)$\times$SO(2). These will be the isometry groups of the fiber metric; the fact that the $S^4$ is fibred over $\Sigma_3$ will break the isometry group further, down to a flavor SU(2) in the ${\cal N}=1$ case and down to SO(2) (which is then the R-symmetry group) in the ${\cal N}=2$ case.\footnote{If $E$ is the total space of an $F$-fibration over a base space $B$, the isometries of $B$ are promoted to isometries of $E$, but often the isometries of $F$ are not. To see this, write the metric on $E$ as $ds^2_E= g_{ij}^F Dx^i Dx^j + g_B$, where $x^i$ and $g_{ij}^F$ are the coordinates and metric on $F$, and $Dx^i\equiv dx^i + A^i$; $A^i$ is a connection on $B$, which takes values in the space of isometries of $F$. Now it can be shown that an isometry $\xi$ of $F$ preserves the total metric $g^E$ if and only if $d\xi + [\xi,A]=0$, where the bracket is the Lie bracket of vectors on $F$; in other words, if $\xi$ is a covariantly constant section of the bundle ${\rm ad}(E)$, the adjoint bundle associated to $E$. If $F=S^1$, the Lie bracket vanishes and one can take $\xi$ to be constant over $B$. With more complicated $F$'s, ${\rm ad}(E)$ is often non-trivial and does not have a non-trivial global section; thus $\xi$ cannot be promoted to an isometry of $E$.} 

Geometrically, the ${\cal N}=1$ solution can be thought of as arising when one wraps an M5 stack on a submanifold $\rr^3 \times \Sigma_3 \subset \rr^4 \times$ a $G_2$ manifold; supersymmetry demands $\Sigma_3$ to be an ``associative'' submanifold. In this case, four of the five scalars transverse to the M5 span directions in the $G_2$ manifold, corresponding to the SO(4); these scalars will be ``twisted'', meaning that they are really sections of the normal bundle. The remaining scalar represents the transverse direction inside the $\rr^4$, and is not fibred. The ${\cal N}=2$ solution, on the other hand, arises when wrapping an M5 stack on a submanifold $\rr^3\times \Sigma_3 \subset \rr^5\times$ Calabi--Yau$_6$; supersymmetry demands $\Sigma_3$ to be a ``special Lagrangian'' submanifold. In this case, three scalars are inside the CY$_6$, and two trivial ones are in the flat directions; this corresponds to the SO(2)$\times$SO(3).

Accordingly, there are two different coordinate systems on $S^4$ that are appropriate to describe these two cases. 

For the $\nn=1$ compactification, it is convenient to write the $S^4$ as:
\begin{equation} \label{s4n1}
ds^2_{S^4} = d\alpha^2 + \sin^2\alpha ds^2_{S^3} \ .
\end{equation}
The metric on the $S^3$ can be written in terms of the Maurer--Cartan forms as $ds^2_{S^3} = \frac{1}{4}\sigma^i \sigma^i$, with $d\sigma^i = \frac{1}{2}\epsilon^{ijk} \sigma^{jk}$. Alternatively we can choose Hopf coordinates which are appropriate to study the reduction to ten dimensions:
\begin{equation}
	ds^2_{S^3} = \frac{1}{4} ds^2_{S^2} + (d\beta + A)^2\ ,
\end{equation}
where $dA=-\frac{1}{2} {\rm vol}_{S^2}$. The transformation rules between these two sets of coordinates is given in detail in appendix \ref{s3s2}.

After wrapping the M5 on $\Sigma_3$, which corresponds to replacing AdS$_7$ with AdS$_4 \times \Sigma_3$, the metric of the $S^4$ will be deformed in such a way that the original SO(5) isometry will be broken to the subgroups mentioned above. Part of the residual symmetry gets mixed with the local Lorentz group of the three manifold where the M5 is wrapped, meaning that a subspace of $S^4$ which is left untouched by the supersymmetric deformation gets fibered over $\Sigma_3$.

In the $\nn=1$ case, the $S^4$ metric \eqref{s4n1} gets deformed in such a way as to preserve the shape of the $S^3$:
\begin{equation} \label{s4n1def}
ds^2 \left( S^4_{\nn=1} \right) = d\alpha^2 + \frac{\sin^2 \alpha}{w} Ds^2_{S^3} \ , \qquad w = \frac{5 + 3 \cos^2 \alpha}{8} \ .
\end{equation}
Notice that the supersymmetric deformations are encoded into a single ``distortion'' function $w$. The upper case on $Ds^2_{S^3}$ means that the $S^3$ is now fibred over $\Sigma_3$. In terms of the Maurer--Cartan forms:
\begin{equation}
	Ds^2_{S^3} = \frac{1}{4} \mu^i \mu^i\ ,
\end{equation}
where $\mu^i = \sigma^i - \omega^i$, and the $\omega^i$ are related to the spin connection on the base space $\Sigma_3$:
\begin{equation}\label{eq:omegai}
	\omega^i =\frac{1}{2}\epsilon^{ijk} \omega^{jk} \ .
\end{equation} 
Alternatively, we can switch to Hopf coordinates and write $Ds^2_{3}= D\beta^{2} + \frac{1}{4} Ds^2_{S^2}$, where
\begin{equation}\label{eq:Ds2}
	 Ds^2_{S^2} \equiv Dy^i Dy^i\ ,
\end{equation}
and 
\begin{equation}\label{eq:y}
	y^i = (\sin \theta \cos \varphi, \sin \theta \sin \varphi, \cos \theta)
\end{equation}
are the $\ell=1$ spherical harmonics on $S^2$, which can be thought of as ``constrained coordinates'': $y^i y^i=1$.  Their covariant derivatives are defined as
\begin{equation}\label{eq:Dy}
	Dy^i \equiv dy^i + \epsilon^{ijk} y^j \omega^k \ ,\qquad D\beta \equiv d\beta + A - \frac{1}{2} y^k \omega^k\ ,
\end{equation}
as explained in appendix \ref{s3s2}. The complete metric describing the $\nn=1$ twist can finally be rewritten in a very compact form in terms of the function $w$ introduced in \eqref{s4n1def}:
\begin{equation}\label{ads7s4n1}
m^2 ds^2_{11,\ \nn=1}  = w^{1/3} \left[ ds^2_{\rm AdS_4} + \frac{4}{5} ds^2_{\Sigma_3} + \frac{2}{5} ds^2 \left( S^4_{\nn=1} \right) \right] \ ,
\end{equation}
where the three manifold $\Sigma_3$ is constrained by supersymmetry to be a (compact quotient of) a maximally symmetric space of negative curvature, with Ricci scalar $R$ normalized to $-6$. (The constant $m$ will be fixed in the next subsection.)

The $\nn=2$ compactifications can be studied using ``topological joint'' coordinates on the $S^4$:
\begin{equation}
ds^2_{S^4} = d\alpha^2 + \sin^2 \alpha\ d\beta^2 + \cos^2 \alpha\ ds^2_{S^2} \ ,
\end{equation}
Morally, $\beta$ is the angular coordinate inside the two transverse directions inside the $\rr^5$, while the $S^2$ are the angular directions inside the three transverse directions inside the CY$_6$. In this case the twisting amounts to fibering the $S^2$ over $\Sigma_3$ and the resulting supersymmetric deformation of the $S^4$ metric takes the form:
\begin{equation} \label{s4n2def}
ds^2\left( S^4_{\nn=2} \right)  = d\alpha^2 + \frac{\sin^2 \alpha}{2 w_2}\ d\beta^2 + \frac{\cos^2 \alpha}{4 w_2}\ Ds^2_{S^2} \ , \qquad w_2 = \frac{1 + \sin^2 \alpha}{2} \ .
\end{equation}
The $S^2$ is fibered over $\Sigma_3$ according to (\ref{eq:Ds2}), (\ref{eq:Dy}). The complete eleven-dimensional metric can again be expressed nicely in terms of the warping function $w_2$ that measures the deformation of the $S^4$ metric in this coordinate system. We get\footnote{Notice that we adopted a slightly different normalization with respect to  \cite{gauntlett-macconamnha-mateos-waldram}, which amounts to choosing the Ricci scalar to be $R=-6$ and the integration constant $\beta = 1/2$. Our normalization allows to get the same radius for AdS$_4$ and $\Sigma_3$, which is indeed the case for the original solution found in 7d maximal gauged supergravity\cite{pernici-sezgin}.}
\begin{equation}
m^2 ds^2_{11,\ \nn=2}  = w_2^{1/3}  \left[ ds^2_{\rm AdS_4} + ds^2_{\Sigma_3} + \frac{1}{2} ds^2 \left( S^4_{\nn=2} \right) \right] \ .
\end{equation}
Again, a supersymmetric solution exist only for $\Sigma_3$ of negative curvature. 

This solution is less interesting for our purposes, since it cannot be reduced to ten dimensions without breaking supersymmetry. Indeed, given that $\del_{\alpha}$ is not an isometry and that we also want to preserve the twisted $S^2$ factor in (\ref{s4n2def}) as it is, the only possibility would be reducing along the $\beta$ coordinate. However in this case the U(1) transformation $\beta \rightarrow \beta + \delta \beta$ coincides with the R-symmetry of the solution, meaning that all the components of the Killing spinor $\eta$ will depend on this coordinate. Hence imposing the condition $\del_{\beta} \eta = 0$ would break all supersymmetry.

On the other hand, the ${\cal N}=1$ solution (\ref{ads7s4n1}) has no R-symmetry so we can reduce it to ten dimensions along the $\beta$ direction without breaking any further supersymmetry.

\subsection{Compactification from IIA supergravity}\label{sub:IIA}

The metric for a warped AdS$_7$ solution in ten dimensions reads
\begin{equation} \label{ads7m3}
ds^2_{10}=e^{2A}ds^2_{\rm AdS_7} + ds^2_{M_3} .
\end{equation}
Supersymmetry requires the presence of SU(2) R-symmetry; this implies that $M_3$ must contain an $S^2$. It was shown in \cite{afrt} that indeed $M_3$ is an $S^2$-fibration over an interval:
\begin{equation} \label{m3}
ds^2_{M_3}=dr^2 + \frac{(1-x^2)}{16}e^{2A}ds^2_{S^2} \ .
\end{equation}
Here, $x$ and $A$ are functions of the coordinate $r$; at the extrema of the interval we have $x\to \pm 1$, so that topologically $M_3\cong S^3$. Notice that $r$ is different from the coordinate $\alpha$ we used in eleven dimensions; of course the two are related by a radial diffeomorphism.

The simplest solution within this class is the massless one, which of course corresponds to the reduction to ten dimensions of the AdS$_7 \times S^4$ background (\ref{ads7s4}) and can be given analytically as: 
\begin{equation} \label{ads7massless10d}
x=\cos \alpha \ , \qquad e^{2A} = \frac{R^3}{2}\sin\alpha \ , \qquad e^{2\phi}=\frac{R^3}{8}  \sin^3\alpha \ .
\end{equation}
A careful analysis of the ten-dimensional geometry reveals that we are in presence of a D6 (anti-D6) singularity at the two poles. This analysis was done in \cite{afrt}, where it is also shown how to reduce from eleven to ten dimensions along the Hopf fiber parametrized by $\beta$ in (\ref{s4n1}). Supersymmetry is partially preserved imposing the condition $\partial_{\beta} \eta = 0$ on the $S^4$ Killing spinor, 
which amounts to projecting out half of its components.

It is now crucial to notice that the same coordinate system is also appropriate to describe the reduction of the $\nn=1$ AdS$_4$ background (\ref{ads7s4n1}), which ends up being an $\nn=1$ solution in ten dimensions as well. The resulting ten-dimensional metric can be written in the following form:
\begin{equation} \label{433spontaneous}
ds^2_{10}  = \left( \frac{8}{5} \right)^{3/2}  e^{2A} \left( ds^2_{\rm AdS_4} + \frac{4}{5} ds^2_{\Sigma_3} \right)  + ds^2_{M_3} \ ,
\end{equation}
where the internal space metric can be expressed as in (\ref{s4n1def}) in terms of a deformation function $w$ and reads:
\begin{equation} \label{eq:ds2M3}
ds^2_{M_3}= \left( \frac85 \right)^{1/2} \left( dr^2 + \frac{1-x^2}{16 w} e^{2A}Ds^2_{S^2} \right) \ , \qquad w = \frac{5 + 3 x^2}{8} \ ,
\end{equation}
where the radial coordinate $r$ and the two functions $A, x$ entering the last two formulas were defined in AdS$_7$ by (\ref{ads7massless10d}). 
The choice to express the AdS$_4$ metric (\ref{433spontaneous}, \ref{eq:ds2M3}) in terms of the quantities entering the AdS$_7$ metric (\ref{ads7m3}, \ref{m3}) is in order to highlight the similarity between the two formulas, and turned out to be the key ingredient in the formulation of the universal map described in section (\ref{74sec}). In writing (\ref{eq:ds2M3}), we have also expressed the constant $m$ we had in (\ref{ads7s4n1}) in terms of the AdS$_7$ radius as $m^3 R^3 = \left(\frac{8}{5}\right)^2$; this will be convenient for flux quantization, to be discussed later in section (\ref{fluxquant}).

 
(\ref{433spontaneous}) provides a first example of $\nn=1$ AdS$_4$ solution in type IIA supergravity that can be interpreted as compactification of an AdS$_7$ solution --- namely of the massless reduction to IIA of AdS$_7\times S^4$, which was worked out in \cite[Sec.~5.1]{afrt}.

The goal of this paper is finding more general solutions adding a massive perturbation $F_0$ to this background, solutions which would in turn correspond to compactifications of the massive AdS$_7$ solutions (\ref{ads7m3}). Our Ansatz for the metric will consist in keeping the same terms as in (\ref{433spontaneous}) and (\ref{eq:ds2M3}), but with different functions $f=f(r)$, $g=g(r)$:
\begin{equation} \label{433metric}
ds^2_{10}=e^{2A}ds^2_{\rm AdS_4} +g^2 ds^2_{\Sigma_3}  + dr^2 + f^2 Ds^2_{S^2}.
\end{equation}
In other words we assume the metric to be invariant under a simultaneous SO(3) local Lorentz transformation on $\Sigma_3$, and an identical SO(3) rotation acting on the $S^2$. This ``diagonal'' SO(3) acts on the vielbein $e^i$ of $\Sigma_3$ and on the $y^i$ in (\ref{eq:y})
\begin{equation}\label{eq:SO3D}
	e^i \rightarrow O^{ij} e^j \ ,\qquad y^i \rightarrow O^{ij} y^j\ . \qquad ({\rm SO}(3)_{\rm D})
\end{equation}
This ``twisted symmetry'' will play a crucial role in formulating our Ansatz for supersymmetry in the next section.

\section{Technology for AdS$_7$ to AdS$_4$ compactifications}\label{sec:tech}

As anticipated in the introduction, we will deal with supersymmetry using generalized complex geometry techniques. These allow to reformulate all the data of a given vacuum into a pair of polyforms on the internal space $M_6$, the so-called pure spinors $\Phi_{\pm}$. In other words, generalized complex geometry provides a way of getting rid of the spinors and rewriting the supersymmetry conditions only in terms of forms, which are simpler to handle. 

Nevertheless we still have to formulate a good Ansatz for the pure spinors $\Phi_\pm$. This will have to reflect that for us $M_6$ will be a fibration of $M_3$ over $\Sigma_3$: 
\begin{equation}
	\xymatrix{ 
	M_3\ \ar@{^{(}->}[r] & M_6\ar[d]\\
	& \Sigma_3\ .}
\end{equation}

To warm up, we will analyze the supersymmetric spinors for the solutions reviewed in the previous section.


\subsection{Twisted Spinors}\label{sub:twsp}

Let us start by looking at the supercharges in AdS$_7$ for the maximally supersymmetric eleven-dimensional background (\ref{ads7s4}). They can be written as: 
\begin{equation}
\epsilon^{11} = \sum\limits_{a=1}^4 \zeta^a \otimes \eta^a + c.c. \ ,
\end{equation}
where $\zeta$ is the Killing spinor on AdS$_7$ and $\eta$ the one on $S^4$, the corresponding gamma representation being:
$\Gamma_{\mu}^{(7+4)} = \gamma_{\mu}^{(7)} \otimes \gamma \ , \ \Gamma_{m+6}^{(7+4)} = 1 \otimes \gamma^m \ .$

Let us first focus on $\eta$, which in our coordinate system (\ref{s4n1}) reads \cite[App.~B]{afrt} 
\begin{equation} \label{s4ks}
\eta_{S^4}= \rm{exp}\left[{\frac{\alpha}{2} \gamma \gamma_1}\right] \rm{exp}\left[{\frac{\theta}{4}\gamma_{12}+\frac{\theta - \pi}{4}\gamma_{34}}\right] \rm{exp}\left[ \frac{\beta+\varphi}{4}\gamma_{13} + \frac{\beta -\varphi }{4}\gamma_{24} \right] \eta_0 \ .
\end{equation} 
$\theta$ and $\varphi$ are the coordinates on $S^2$, and $\beta$ parametrizes the Hopf fiber in (\ref{s4n1}); $\eta_0$ is a constant spinor. In order to reduce this spinor to ten dimensions along the $\beta$ direction, we have to impose the condition $\partial_{\beta} \eta = 0$, which is easily achieved imposing the projection $(\gamma_{13}+\gamma_{24}) \eta_0=0$, which is equivalent to
$\gamma\eta_0 = - \eta_0$. This projection keeps only half of the components, those with negative chirality, so that the solution is half BPS in ten dimensions. 

We now choose the following decomposition for the 4d gamma matrices: $\gamma^i = \hat\sigma^i \otimes \sigma^1 , \gamma^4 = 1 \otimes \sigma^3$, where $\hat\sigma^i = \{ \sigma^3 , \sigma^1, \sigma^2 \}$ and $\sigma^i$ are the Pauli matrices. The condition $\gamma\eta_0 = - \eta_0$ is easily solved by $\eta_0 = (\chi_0\ , - i \chi_0)$. With some more effort, the full $S^4$ Killing spinor (\ref{s4ks}) turns out to admit a natural decomposition in terms of an $S^2$ Killing spinor:
\begin{equation} \label{s4s2ks}
\eta_{S^4} = \left( \begin{array}{c}  e^{-\frac{i\delta_1}{2} \sigma^3} \chi_{S^2}  \\ i e^{\frac{i\delta_2}{2} \sigma^3} \chi_{S^2} \end{array} \right) \ .
\end{equation}
The $S^2$ Killing spinor can be written explicitly as $\chi_{S^2} = \exp \left[\frac{i\theta }{2} \sigma^1\right] \exp \left[\frac{\varphi}{2} \sigma^{12} \right] \tilde{\chi}_0$, for a new constant spinor which is related to the old one by a simple unitary transformation: $\tilde{\chi}_0= \frac{1}{2}(1-i\sigma^1)(1+i\sigma^3) \chi_0$. After the reduction to ten dimensions, the spinor dependence on the coordinate $\alpha$ gets factorized in an overall unitary transformation, which is parametrized by two angles that are related to $\alpha$ by: $\delta_1 = \alpha + \frac{\pi}{2} \ , \delta_2 = \delta_1 + \pi$.

The gamma matrix representation we have chosen is already appropriate for the reduction from eleven to ten dimensions.  Indeed, chirality in ten dimensions is given by the eigenvalues of $\gamma^4$, which in our basis is $\gamma^4 = 1 \oplus -1$. The spinor $\eta$ decomposes as $(\chi^1 ,\chi^2)$, or equivalently as $\eta=\chi^1 \otimes v_+ + \chi^2 \otimes v_-$, where $v_{\pm}$ are $\sigma^3$  eigenvectors and the two spinors on $M_3$ are given by
\begin{equation}\label{eq:chi120}
	\chi^1 = e^{-\frac{i\delta_1}{2} \sigma^3} \chi_{S^2} \ ,\qquad \chi^2 = i e^{\frac{i\delta_2}{2} \sigma^3} \chi_{S^2} \ .
\end{equation}
We end up with the following two supercharges with opposite chirality in type IIA supergravity:
\begin{equation} \label{ads7supercharges}
\epsilon^1_+ = \zeta \ \chi^1 v_+ + {\rm c.c.} \ ,\qquad
\epsilon^2_- = \zeta\ \chi^2 v_- + {\rm c.c.} \ .
\end{equation}
The corresponding gamma matrices representation is given by:
$\Gamma_{\mu}^{(7+3)} = \gamma_{\mu}^{(7)} \otimes 1 \otimes \sigma^2 \ , \ \Gamma_{i+6}^{(7+3)} = 1 \otimes \hat\sigma^i \otimes \sigma^1 \ , \ \Gamma^{(7+3)} = 1 \otimes 1 \otimes \sigma^3$.

(\ref{ads7supercharges}) is also the spinor decomposition given in \cite{afrt} for the AdS$_7 \times M^3$ solutions in massive type IIA supergravity. The SU(2) isometry of the $S^2$ is preserved by the massive deformations, and it is in fact the R-symmetry of the solution. This is implemented by having ${\zeta \choose \zeta^c}$ transform as a doublet, and at the same time the internal spinors:
\begin{equation} \label{chidoublets}
\chi^{1a} \equiv \left( \begin{array}{c}  \chi^1 \\ \chi^{1c} \end{array} \right) \in 2 \ , \qquad \chi^{2a} \equiv \left( \begin{array}{c}  \chi^2 \\  -\chi^{2c} \end{array} \right) \in 2 \ .
\end{equation}
This is indeed also the case for the massless case (\ref{ads7supercharges}) we just discussed. This was initially not assumed in \cite{afrt}, but it is indeed a consequence of supersymmetry, as can be checked from the bispinors given there.\footnote{The spinors for the AdS$_7$ solutions in \cite{afrt} were obtained by I.~Bakhmatov in unpublished work. We thank him for sharing his work with us.} In other words, massive deformations do not alter the transformation properties of $\chi^1$ and $\chi^2$ under SU(2). One can actually even check from \cite[Eq.(4.23)]{afrt} that the spinors $\chi^1$ and $\chi^2$ of the massive AdS$_7$ solutions are given again by (\ref{eq:chi120}), replacing however $\alpha$ with $\psi \equiv \arccos x$, an angle whose role will become clear later in this section. 

We now want to further decompose the AdS$_7$ spinor in a way which is appropriate to describe an AdS$_4$ compactification. This is easily accomplished with $\zeta_{\rm AdS_7} \rightarrow \zeta_{\rm AdS_4} \otimes \tilde{\chi}$, where $\tilde{\chi}$ is a complex spinor on the three manifold $\Sigma_3$ and the Killing spinor on AdS$_4$ is a real non chiral spinor that we can write as: $\zeta_{\rm AdS_4} = \zeta + \zeta^*$. 
The corresponding gamma matrices decomposition is: $\gamma_{\mu}^{(7)} = \gamma_{\mu}^{(4)} \otimes 1 \ , \gamma_{i+3}^{(7)} = \gamma^{(4)} \otimes \tilde{ \sigma}^i$, with charge conjugation matrix $B^{(7)} = 1 \otimes i \sigma^2$.

If we now plug this decomposition into the ten dimensional gamma matrices we immediately realize that a change of basis is needed in order to get a proper $10=4+6$ representation. 
We thus rotate the ten-dimensional spinors accordingly to $\epsilon \rightarrow O \epsilon$, where the change of basis is parametrized by a matrix of the form: $O = \frac{1}{\sqrt{2}}(1 + i \rho)$, where $\rho^2 = 1$ in such a way that $O^{-1} = O^* = \frac{1}{\sqrt{2}}(1 - i \rho) $. The corresponding transformation law for the gamma matrices is $\Gamma \rightarrow O \Gamma O^{-1}$, which amounts to: $\Gamma \rightarrow \Gamma$ if $\Gamma$ and $\rho$ commute, and to: $\Gamma \rightarrow i \rho \Gamma$ if $\Gamma$ and $\rho$ anticommute. The charge conjugation matrix transforms as $B \rightarrow O B (O^*)^{-1}$. 

A proper choice is $\rho = \gamma^{(4)} \otimes 1 \otimes 1 \otimes \sigma^2$, which leads to our final $4+3+3$ gamma matrices representation:
\begin{align} \label{433gamma}
\Gamma_{\mu}^{(4+3+3)} & = i \gamma ^{(4)} \gamma_{\mu}^{(4)} \otimes 1 \otimes 1 \otimes 1 \ , \nonumber \\
\Gamma_{i+3}^{(4+3+3)} & =  \gamma ^{(4)} \otimes \tilde{\sigma}^i \otimes 1 \otimes \sigma^2 \ , \\
\Gamma_{i+6}^{(4+3+3)} & = \gamma ^{(4)}  \otimes 1 \otimes \hat\sigma^i \otimes \sigma^3 \ ,
\nonumber 
\end{align}
where the index $i=\{1,2,3\}$ runs over both the manifold $\Sigma_3$ where the branes are wrapped and on $M_3$.
In this basis chirality and charge conjugation are represented as:
\\ $\ \Gamma = \gamma ^{(4)}  \otimes 1 \otimes 1 \otimes (-\sigma^1) \ , \  B=1  \otimes i \sigma^2 \otimes i \sigma^2 \otimes \sigma^3$.

The resulting transformed supercharges are:
\begin{equation} \label{epsilon12}
\epsilon^1_+ = \zeta_+  (\tilde{\chi} \chi^1 +\tilde{\chi}^c \chi^{1c} ) w_+ + {\rm c.c.} \ ,\qquad \epsilon^2_- = \zeta_+ (\tilde{\chi} \chi^2 -\tilde{\chi}^c \chi^{2c} )w_-+ {\rm c.c.}\ , 
\end{equation}
where $w_{\pm}$ are eigenvectors of $-\sigma_1$, namely $w_{\pm}= \frac{1}{\sqrt{2}}(v^+ \mp v^-)$. 

As we anticipated in the previous section, it is very convenient to rewrite our spinor Ansatz in such a way as to make the twisted symmetry manifest. 
We already know from (\ref{chidoublets}) the transformation properties for $\chi^1$ and $\chi^2$, and we also know that the AdS$_4$ spinor $\zeta$  has to be invariant. Therefore it looks natural to assume that the spinor $\tilde{\chi}$ living on $\Sigma_3$ transforms under local Lorentz transformation on $\Sigma_3$ in such a way as to compensate the variation of $\chi^a$ under $S^2$ isometry. This is the analogue in our case of the discussion about wrapped M5-branes at the beginning of section \ref{cft6sec}, except that of course our solutions will not originate from wrapped M5s, but morally from wrapping the NS5--D6--D8 systems of \cite{hanany-zaffaroni-6d,brunner-karch}.
 
We can thus introduce a new SU(2) doublet $\tilde{\chi}^a \equiv \left( \begin{smallmatrix} \tilde{\chi} \\ \tilde{\chi}^c \end{smallmatrix} \right)$ transforming in the $\bar{2}$.\footnote{Of course the $SU(2)$ representations $2$ and its conjugate $\bar{2}$ are equivalent. What we want to highlight here is that if $\chi^a$ transform as $\chi^a \rightarrow U^{ab} \chi^b$ then $\tilde{\chi}^a$ has to transform as $\tilde{\chi}^a \rightarrow U^{*ab} \tilde{\chi^b}$ in such a way to make the product $\tilde{\chi}^a \chi^a$ invariant.}
At this point it is crucial to notice that both $\tilde{\chi}$ and its conjugate carry a spacetime spinor index $\alpha$, or in other words assuming that they also transform as a doublet amounts to imposing a condition on them, the twisting condition
\begin{equation}\label{eq:twcond}
U^{*ab} \tilde{\chi}^{ b \alpha} = \tilde{\chi}^{ a \beta} U^{\beta \alpha} \ ,
\end{equation}
where we have introduced a spinor index $\alpha=\{1,2\}$ on $\Sigma_3$. This constraint is solved by choosing the twisted spinor to be equal to the epsilon tensor:\footnote{We can be a bit more explicit by choosing a representation for the gamma matrices on the tangent space to $\Sigma_3$ to be: $\tilde{\sigma}^i = - (\sigma^{i})^*$. We then define the spinor rotation matrix  with respect to the euclidean rotation matrix in the following way:
\begin{equation}
O^{ij}\sigma^j = U^{\dagger} \sigma^i U \ , \qquad O^{ij}\tilde{\sigma}^j = U^{T} \tilde{\sigma}^i U^* \ .
\end{equation} 
This identity implies that the spinor $\tilde{\chi}$, which is defined with respect to the generators in the $\tilde{\sigma}^i$ representation, transforms under local Lorentz transformation as $\tilde{\chi}^{\alpha} \rightarrow (U^{T})^{\alpha \beta} \tilde{\chi}^{\beta}$. We also want $\tilde{\chi}$ and its conjugate to transform as a doublet under the same symmetry: $\tilde{\chi}^a \rightarrow U^{*ab} \tilde{\chi}^b$. Then the full spinor along $\Sigma_3$ is represented by a $2 \times 2$ matrix $\tilde{\chi}^{ a \alpha}$, which gets constrained  by setting the two transformation laws to be equivalent; this leads to (\ref{eq:twcond}).}
\begin{equation}\label{eq:chiaaeps}
\tilde{\chi}^{ a \alpha} = \epsilon^{ a \alpha} = \left( \begin{array}{c c}  0 & 1 \\ -1 & 0 \end{array} \right) .
\end{equation}
This is indeed the explicit form for the twisted spinor which is given in the $AdS_4 \times \Sigma_3$ solutions of seven-dimensional gauged supergravity in \cite{pernici-sezgin}.

To summarize this long discussion about spinors, we achieved the goal of rewriting the two six dimensional internal spinors in a form which is  manifestly invariant under the twisted SU(2) symmetry: 
\begin{equation} \label{M6spinors}
\eta^1_+=\tilde{\chi}^a \chi^{1a}w_+\ , \qquad \eta^2_- = \tilde{\chi}^a \chi^{2a} w_- \ .
\end{equation}
This is also consistent with the four dimensional spinor $\zeta$ being a singlet. We finally have a good SU(2)$_{\rm D}$ invariant spinor Ansatz and we can proceed in our analysis of the internal space structure. 



\subsection{Twisted Forms} \label{twistedforms}

We now discuss the fibration of $M_3$ over $\Sigma_3$ from the point of view of forms. We will focus our attention on those that are invariant under the diagonal SO(3)$_{\rm D}$ in (\ref{eq:SO3D}).
As we mentioned there, this action is a symmetry of the metric (\ref{433metric}), whose internal part we repeat here:
\begin{equation}
ds^2_{M_6} = g^2 e^i e^i +  dr^2 + f^2 Ds^2_{S^2}\ ,
\end{equation}
where $Ds^2_{S^2}$ is the fibred $S^2$ metric defined in (\ref{eq:Ds2}), (\ref{eq:Dy}), and we have written the metric on $\Sigma_3$ in terms of its vielbein $\{e^i\}$; their Cartan structure equation reads, in terms of (\ref{eq:omegai}), $de^i = \epsilon^{ijk} e^j \omega^k$. We also demand
\begin{equation}\label{eq:ms}
	R^{ij} = \frac{R}{6} e^i\wedge e^j\ ,
\end{equation}
where $R^{ij} = \frac{1}{2} R^{i j}{}_{\mu \nu} dx^{\mu \nu}$; this appears in the derivative of the spin connection $d\omega^i = \frac{1}{2} \epsilon^{ijk} (\omega^{jk} + R^{jk})$. (\ref{eq:ms}) is valid on a compact quotient of a maximally symmetric space. 

We can construct only two SO(3)$_{\rm D}$ invariant one-forms: $\{ dr , y^i e^i \}$. A third possible candidate is vanishing: $y^i Dy^i = 0$. It is easy to see that $d(y^ie^i) = Dy^i e^i$, which suggests that the subspace of invariant forms is closed under derivation, which will indeed turn out to be true.

Moving on to two-forms the structure becomes richer as there are five SO(3)$_{\rm D}$ invariant combinations living on $M_6$ that are given by:\footnote{We chose the notation $\omega_i$, as is relatively standard for a basis of two-forms. These should not be confused with the spin connection.}
\begin{equation} \label{wi}
\begin{split}
	&\omega_1 = \frac{1}{2} \epsilon^{ijk}y^i Dy^{jk} \ ,\qquad
	\ \omega_2 = e^i Dy^i \ ,\qquad
	\ \omega_3 =  dr y^i e^i \ ,\\ 
	&\ \omega_4 = \epsilon^{jik} e^i y^j Dy^k \ ,\qquad
	\ \omega_5 =  \frac{1}{2} \epsilon^{ijk}y^i  e^{jk}\ .
\end{split}
\end{equation}
(Notice that in this section we will omit wedge products to make the expressions more readable.) Their exterior derivatives read:
\begin{equation} \label{formsderivatives}
\begin{split}
	&d \omega_1 =  -\frac{R}{6} y^i e^i \omega_4 \ ,\qquad
	\ d \omega_2 = 0 \ ,\qquad
	\ d \omega_3 = dr \omega_2 \ , \\  
	&\ d \omega_4 = 2 y^i e^i  \left( \omega_1 - \frac{R}{6} \omega_5 \right) \ ,\qquad
	\ d \omega_5 = y^i e^i  \omega_4 \ .	
\end{split}
\end{equation}
In the space spanned by these invariant two-forms $\omega_i$, there is only one closed two-form which is not exact:
\begin{equation}\label{eq:o15}
	\omega_1 + \frac{R}{6} \omega_5\ .
\end{equation}
This will be relevant in a later discussion regarding the flux quantization.

We now consider four-forms. It is natural to define
\begin{equation}
	\omega_{AB} \equiv \omega_A \wedge \omega_B\ .
\end{equation}
Only five of these are non-vanishing, corresponding to the Hodge duals of the two-forms $\omega_i$:
\begin{equation}
\begin{split}
	&\star \omega_1 = f^2g^3\omega_{15}  \ , \qquad
	\ \star \omega_2 = -2f^2g^3 \omega_{23} \ , \qquad
	\ \star \omega_3  =f^2g^3 \omega_{15}\ , \\
	&\ \star \omega_4  = -2f^2g^3\omega_{43} \ , \qquad 
	\ \star \omega_5  = f^2g^3\omega_{35} \ ,
\end{split}
\end{equation}
where the Hodge star on $M_6$ is computed with respect to the volume form ${\rm vol}_6 = f^2 g^3 \omega_1 \wedge \omega_3 \wedge \omega_5$, which is of course the only non vanishing six form.

Finally it is worth noticing that:
\begin{equation}\label{eq:omsquare}
\omega_{22} = \omega_{44} = -2 \omega_{15} \ ;
\end{equation}
this implies that we have a triplet of two-forms $\{ \omega_2 ,\ \omega_4, \  \omega_1 - \omega_5 \}$ that square to the same four-form and that are orthogonal to each other. This is exactly the set of algebraic constraints that define a so called SU(2) structure on $M_6$. This will be useful in defining the pure spinors in the section \ref{3dps}.

\subsection{Pure spinors and supersymmetry}

We will now give a quick review of the essentials of the pure spinor formalism, which will allow us to formulate supersymmetry in a very compact fashion. For more details see for example \cite{gmpt3}.

A warped AdS$_4$ compactification is a spacetime of the form
\begin{equation}
ds^2_{10}=e^{2A}ds^2_{\rm AdS_4} + ds^2_6 \ ,
\end{equation}
where $ds^2_6$ is the metric on the internal space $M_6$, and $A$ is a function of $M_6$ called \textit{warping}. The BPS equations for a string vacuum with this geometry can be rewritten \cite{gmpt2} using the language of generalized geometry in terms of the so called \textit{pure spinors}, a pair of polyforms on the internal space. If we take the standard $10=6+4$ decomposition for the supersymmetry parameters:
\begin{equation} \label{46spinor}
\epsilon^1 = \zeta_+ \eta^{1}_+ + c.c. \ ,\qquad \epsilon^2 = \zeta_+ \eta^{2}_- + c.c\ , 
\end{equation}  
we can define the pure spinors $\Phi_{\pm}$ in terms of the internal parameters as:
\begin{equation}\label{eq:Phipm}
\Phi_{-} \equiv \eta^1_+ \otimes \left( \eta^2_- \right)^{\dagger} \ , \qquad  \Phi_{+} \equiv \eta^1_+ \otimes \left( \eta^{2c}_+ \right)^{\dagger} \ ,
\end{equation}
where $\pm$ denotes even/odd forms. In the generic case the pure spinors can be written in terms of a so called SU(2) structure on $M_6$, given by a complex one-form $z$, a complex two-form $\omega$ and a real two-form $j$, such that
\begin{equation}\label{eq:SU2}
	\omega \wedge \bar{\omega} = j^2 \ ,\qquad \omega^2 = 0 \ .
\end{equation}
The parametrization is:
\begin{equation} \label{dielectric}
e^{-b } \Phi_+ = \rho\ e^{i \theta} e^{-i J_{\psi}} \ , \qquad e^{-b }  \Phi_- = \rho \tan \psi\ z \wedge e^{i \omega_{\psi}} \ ,
\end{equation}
where $\psi$ is the angle between the two spinors $\eta^{1,2}$, and $\rho$ is a real number that determines the norm of the pure spinors. We have also defined the forms:
\begin{equation}\label{eq:Jpsi}
J_{\psi} \equiv \frac{1}{\cos \psi} j + \frac{i}{2} z \wedge \bar{z} \ , \qquad \omega_{\psi} \equiv \frac{1}{\sin \psi} \left( \Re \omega + \frac{i}{\cos \psi} \Im \omega \right) \ , \qquad b=\tan \psi \Im \omega\ ,
\end{equation}
where the real two-form $b$ is called the \textit{intrinsic} $b$-field associated to the pair $\Phi_{\pm}$. One can always obtain a pure spinor pair with vanishing intrinsic $b$ by the action of a so called $b$-transform: 
\begin{equation} \label{btr}
\Phi_{\pm} \rightarrow \Phi_{\pm}^0 = e^{-b \wedge} \Phi_{\pm} \ ,
\end{equation}
which turns out to be a symmetry of the pure spinor equations provided that also the physical NS three-form flux $H$ and the \emph{internal}\footnote{We mean by this the flux with no legs along AdS$_4$; this determines via Hodge duality the external flux, namely the one with legs along AdS$_4$.} RR flux $F = \sum_k F_{2k}$ are transformed to the corresponding auxiliary fluxes given by: 
\begin{equation} \label{btrfluxes}
H^0 = H-d b \ , \qquad F^0 = e^{-b} F.
\end{equation}
We can now write the pure spinor equations \cite{gmpt2,t-reform}:
\begin{equation} \label{pse}
d_{H} \Phi_+ = -2 e^{-A} \Re \Phi_- \ , \qquad {\cal J}_+\cdot d_{H} \left( e^{-3A}\Im \Phi_- \right) = -5 e^{-4A} \Re \Phi_+ + F \ , \qquad d_{H} F = \delta \ ,
\end{equation}
where $d_H \equiv d - H \wedge$ and ${\cal J}_+$ is an algebraic operator associated in a certain way to $\Phi_+$. This operator is reviewed for example \cite{t-reform}, and more concretely in \cite[Sec.~5]{saracco-t}. More specifically, in \cite[Sec.~5.2]{saracco-t}, (\ref{pse}) were analyzed and reduced to the action of a more concrete operator $J_\psi^{-1}\llcorner$, that consists in contracting with the bivector $J_\psi^{-1}$ whose inverse is $J_\psi$ in (\ref{eq:Jpsi}).  This operator is analyzed in detail in appendix \ref{J-1}. It is now easy to see that we can equivalently solve the pure spinor equations (\ref{pse}) for the set of auxiliary fields $\left( \Phi_{\pm}^0,\ F^0,\ H^0 \right)$ and then perform an inverse $b$-transform (\ref{btr}) to get the physical fluxes, as we anticipated.

\subsection{Pure spinors on $\Sigma_3$ and $M_3$} \label{3dps}

We will now focus on the particular case of our interest, namely a $6=3+3$ splitting of the internal space. This requires some extra ingredients of generalized geometry in $d=3$. 

We will give an Ansatz for the bispinors living on the three manifolds $\Sigma_3$ and $M_3$ using the three-dimensional generalized geometry techniques presented in \cite{afrt}. We will later be able to express the full six-dimensional pure spinors in terms of the three-dimensional ones. We already know form the spinor Ansatz (\ref{epsilon12}) that there is a crucial difference between the two three-dimensional factors: namely, we have one single spinor $\tilde{\chi}$ on $\Sigma_3$, while we have two spinors $\chi^1$, $\chi^2$ on $M_3$.  

We start by defining three-dimensional bispinors in a similar way as (\ref{eq:Phipm}) in six dimensions. Namely, on $M$:
\begin{equation}
\psi_1 = \chi_1 \otimes \chi_2^{\dagger} \ ,\qquad\ \psi_2 = \chi_1 \otimes \chi_2^{c \dagger} \ ;
\end{equation}
similarly, on $\Sigma_3$:
\begin{equation}
\tilde{\psi_1} = \tilde{\chi} \otimes \tilde{\chi}^{\dagger} \ ,\qquad \tilde{\psi_2} = \tilde{\chi} \otimes \tilde{\chi}^{c \dagger} \ .
\end{equation}
It is much more convenient to organize the bispinors in the following $2 \times 2$ matrices:
\begin{equation} \label{psi33}
\begin{split}
	\Psi & \equiv 
	\left( \begin{array}{c}
	\chi_1 \\ \chi_1^c 
	\end{array} \right) \otimes \left( \chi_2\ , - \chi_2^c \right)^{\dagger} = 
	\left( \begin{array}{cc}
	\psi_1 &  \psi_2  \\
	-(-)^{\rm deg} (\psi_2)^* & -(-)^{\rm deg} (\psi_1)^*
	\end{array} \right) \ ,  \\
	\tilde{\Psi}  & \equiv 
	\left( \begin{array}{c}
	\tilde{\chi} \\ \tilde{\chi}^c 
	\end{array} \right) \otimes \left( \tilde{\chi}\ ,\ \tilde{\chi}^c \right)^{\dagger} = 
	\left( \begin{array}{cc}
	\tilde{\psi}_1 & \tilde{\psi}_2 \\
	-(-)^{\rm deg}(\tilde{\psi}_2)^* & (-)^{\rm deg}(\tilde{\psi}_1)^* 
	\end{array} \right) \ ,
\end{split}
\end{equation}
where $(-)^{\rm deg}$ acts as $\pm$ on even (odd) forms. 
The advantage of this choice is that now we can expand these $2 \times 2$ hermitean matrices on the two basis $\sigma^{\mu} = (1, \sigma^i )$ and $\tilde{\sigma}^{\mu} = (1,  \tilde{\sigma}^i)$, where $\sigma^i$ and $\tilde{\sigma}^i=-(\sigma^{i})^*$  are the SU(2) generators in the $2$ and $\bar{2}$ representations.

In the case of $\Sigma_3$ we have one single spinor $\tilde{\chi}$, so we can use the expressions
\cite[Eq.(3.14)]{afrt} with $\psi=0,\ \theta_1 =0,\ \theta_2 =0$; the result is:
\begin{equation} \label{Psit}
\tilde{\Psi}_0 = 1 \ ,\qquad\tilde{\Psi}_1 = -  \tilde{e}^i \tilde{\sigma}^i \ .
\end{equation}
The subscript indicates the degree of the forms; we introduced a tetrad $\tilde{e}^i= g e^i$, with $e^i e^i = ds^2_{\Sigma_3}$. The remaining components of $\tilde{\Psi}$ are determined via Hodge duality as $\tilde{\Psi}_2 = - i \star_3 \tilde{\Psi}_1 , \tilde{\Psi}_3 = - i \star_3 \tilde{\Psi}_0 $.  Notice that the expressions (\ref{Psit}) are automatically covariant under the SO(3) of local Lorentz transformations even before solving the supersymmetry equations; with some abuse of language, we will say that they are covariant ``off-shell''. Indeed if we perform a local Lorentz transformation $e^i \rightarrow O^{ij} e^j$, it is clear that this can be traded with $\tilde{\sigma}^i \rightarrow (O^{T})^{ij} \tilde{\sigma}^j = U^* \tilde{\sigma}^i U^{T}$, namely 
the matrix $\tilde{\Psi}$ transforms covariantly as $\tilde{\Psi} \rightarrow U^{*} \tilde{\Psi} U^{T}$.

Things are a bit more complicated on $M_3$ where we have two spinors $\chi_1 , \chi_2$. What happens is that the expression for $\Psi$ is not automatically covariant under the SU(2) that rotates the $S^2$; in \cite{afrt}, it became covariant only ``on shell'', meaning after solving the supersymmetry equations. This in effect means that the analysis there started with random spinors on the $S^2$, and that imposing supersymmetry also required them to be Killing spinors when restricted to the $S^2$. In this paper, we have no need of proving that our solutions are the most general in any sense; so we will just assume the SU(2) covariance from the start. We will simply take the expression for the bispinors \cite[Eq.(4.23)]{afrt} and covariantize it by replacing $d y^i$ with $D y^i = dy + \epsilon^{i j k} y^j \omega^k $, also substituting the AdS$_7$ warping $\frac{1}{4}e^A\sqrt{1-x^2}$ with the AdS$_4$ one $f$. We get:
\begin{equation} \label{Psi}
\Psi_0 = i x \ 1 + \sqrt{1-x^2}\  y^i \sigma^i \ ,\qquad \Psi_1 = \sqrt{1-x^2} dr \ 1 + i \left( x y^i dr + f Dy^i \right) \sigma^i \ .
\end{equation}
Again the remaining components are determined by covariantizing the Hodge duals $\Psi_2 = -i \star_3 \Psi_1 \ ,\ \Psi_3 = -i \star_3 \Psi_0$. 

The matrix $\Psi$ now transforms covariantly under the diagonal symmetry in (\ref{eq:SO3D}), which we can trade for
\begin{equation}
	\sigma^i \rightarrow (O^T)^{ij} \sigma^j = U \sigma^i U^{\dagger}\ .
\end{equation} 
This implies $\Psi \rightarrow U \Psi U^{\dagger}$.

\subsection{Assembling the pure spinors on $\Sigma_3$ and $M_3$} \label{3d3dps}

We will now assemble the pure spinors (\ref{Psi}) and (\ref{Psit}) that we have found on $\Sigma_3$ and $M_3$, and find expression for the six-dimensional pure spinors (\ref{eq:Phipm}).

We start from the odd form $\Phi_-$, that we rewrite as:
\begin{equation} \label{phim1}
\begin{split}
	\Phi_- & = \eta^1_+ \otimes \eta^{2\dagger}_- = \sum\limits_{k=0}^6  \frac{1}{8 k!} \eta^{2\dagger}_-\ \gamma_{M_k \ldots M_1} \ \eta^1_+ dx^{M_1 \ldots M_k} \\ 
	& = \frac{1}{8} \sum\limits_{q=0}^3\sum\limits_{k=0}^3  \frac{1}{q!\ k!} \eta^{2\dagger}_-\ \gamma_{M_q \ldots M_1} \gamma_{\tilde{M}_k \ldots \tilde{M}_1} \ \eta^1_+ dx^{\tilde{M}_1 \ldots \tilde{M}_k} dx^{M_1 \ldots M_q}\ .	
\end{split} 
\end{equation}
We now plug into this formula the spinor Ansatz (\ref{M6spinors}), together with the explicit gamma matrix representation given in (\ref{433gamma}),
and get:\footnote{\eqref{phim2} is obtained after some manipulations that involve computing the quantity $w_- \sigma_3^q \sigma_2^k w_+$ , which is equal to 0 if $q+k$ is even, to 1 if $k$ is even and $q$ is odd, and to $i$ if $k$ is ood and $q$ even.}
\begin{equation} \label{phim2}
\Phi_- = \frac{1}{2} \left[ \left( \tilde{\chi}^b \otimes \tilde{\chi}^{a\dagger} \right)_+ \left( \chi^b_1 \otimes \chi^{a\dagger}_2 \right)_- +
i \left( \tilde{\chi}^b \otimes \tilde{\chi}^{a\dagger} \right)_- \left( \chi^b_1 \otimes \chi^{a\dagger}_2 \right)_+ \right]\ .
\end{equation} 
Comparing this expression with the 3d bispinor matrices we defined in (\ref{psi33}), we realize that we can write more compactly 
\begin{equation}
\Phi_- = \frac{1}{2} {\rm tr} \left( \tilde{\Psi_+^{T}}  \Psi_- + i \tilde{\Psi_-^{T}}  \Psi_+ \right)\ .
\end{equation}
An analoguous expression can be obtained for the even pure spinor $\Phi_+$:
\begin{equation}
\Phi_+ = i\ \eta^1_+ \otimes (\eta^{2c}_+)^{\dagger} = \frac{1}{2} {\rm tr} \left( \tilde{\Psi_-^{T}}  \Psi_- - i\ \tilde{\Psi_+^{T}}  \Psi_+ \right)\ ,
\end{equation}
where the $i$ factor in the definition is chosen in order to get a real zero form part $\Phi_0$. Notice that the pure spinors $\Phi_{\pm}$ are invariant under the twisted symmetry (\ref{eq:SO3D}), as they should be. This can be seen by assembling the transformation rules we found for $\tilde\Psi$ and $\Psi$ in the previous section:
\begin{equation}
\Psi \rightarrow U \Psi U^{\dagger} \ , \ \tilde{\Psi} \rightarrow U^* \tilde{\Psi} U^T \ \Longrightarrow \  \Phi_{\pm}\ \rm{invariant} \ .
\end{equation}

The next step is plugging into $\Phi_{\pm}$ the explicit expressions for the matrices $\Psi^{ab}$ and $\tilde{\Psi}^{ab}$ we gave in (\ref{Psi}) and (\ref{Psit}). As expected, the pure spinors turn out to be naturally expressed in terms of the twisted forms that we introduced in section \ref{twistedforms}. In particular they can be written in the dielectric form (\ref{dielectric}), where 
\begin{equation} \label{zjw}
z =  dr + g\ y^i e^i  \ ,\qquad  j = -fg\ \omega_4 \ ,\qquad \omega= -f g\ \omega_2 + i(f^2 \omega_1 -g^2 \omega_5) \ ;
\end{equation}
from (\ref{eq:omsquare}) we see that this is an SU(2) structure, (\ref{eq:SU2}).\footnote{There are other linear combinations of the $\omega_i$ in (\ref{wi}) that satisfy (\ref{eq:SU2}). One can see that the coefficient of $j$ and $\omega$ along $\omega_3$ has to vanish; the remaining coefficients describe a set of quadratic equations, which can be interpreted as describing a frame $\{{\rm Re} \omega, {\rm Im} \omega, j\}$ in a four-dimensional space of signature $(3,1)$. This might lead to a more general class of solutions, which however would not be interpreted as compactifications of the AdS$_7$ solutions of \cite{afrt}.} We also get for free the identification
\begin{equation}
	x = \cos \psi \ ,
\end{equation}
where $\psi$ is the angle in (\ref{Psi}). This provides a natural interpretation of the variable $x$ in terms of the angle between the two six dimensional spinors $\eta^1$, $\eta^2$. Finally, we also get a vanishing phase $\theta = 0$, which means that we are in the special case considered in \cite[Sec.~5.2]{saracco-t}. The pure spinor equations were analyzed in detail there; (5.16)--(5.18)  in that paper give the constraints on the geometry and the fluxes in terms of the SU(2) structure $(z,j,\omega)$. Recall that before using the equations in that form we have to transform $\Phi_{\pm}$ to the corresponding pair $\Phi_{\pm}^0$ with vanishing intrinsic $b$-field, as in (\ref{btr}). We also have to rescale the pure spinors as $\Phi_{\pm} \rightarrow e^{3A-\phi} \Phi_{\pm}$, which amounts to fixing their norm in (\ref{dielectric}) to $\rho=e^{3A-\phi}\cos \psi$ as in \cite[Eq.~2.2]{saracco-t}. Using the results of section \ref{twistedforms} and appendix \ref{J-1}, after some work the supersymmetry equations reduce to a coupled system of ODE's which we now proceed to give.

\subsection{The system of ODEs} 
\label{sub:ode5}

The result of the analysis of this section is a system of five coupled ODE's in five variables: the three functions in the metric $(f,\ g,\ A)$, the dilaton $\phi$, and the angle between the two six dimensional spinors $x=\cos \psi$. All of these functions depend on the radial coordinate $r$ only. The system reads 
\begin{align} \label{ode5}
\left( \frac{f g^2\ e^{-A}}{\cos \psi} \right)^{'} & = \frac{R f^2 + 6 g^2}{6\ e^A \cos^2 \psi} \ ,  \nonumber \\ 
\left(g\ e^{-A}\right)^{'} & = \frac{e^A(R f^2 + 6 g^2 \sin^2 \psi) - 12 f g^2 \sin \psi}{6 f g\ e^{2A} \cos \psi} \ , \nonumber \\
\left(f\ e^{-\phi}\right)^{'} & = \frac{12 f g^2 (e^A \sin \psi - f)}{e^A(R f^2 -6g^2 \sin^2 \psi) } F_0 \ , \\
\left( g\ e^{3A}\right)^{'} & = \frac{g e^{3A} \cos \psi}{f} + \frac{12 g^3 e^{2A+\phi} (f-e^A \sin \psi) }{(R f^2 -6g^2 \sin^2 \psi)} F_0 \ , \nonumber \\
\left( \frac{g^3 e^{3A}}{f^2} \right)^{'} & =\frac{R}{2} \frac{g e^{3A} \cos \psi}{f} - \frac{2g^3 e^{2A+\phi} \left( 6fg^2 (\cos^2 \psi -3) + e^A \sin \psi (R f^2 +12g^2) \right)  }{ f^2 (R f^2 -6g^2 \sin^2 \psi)} F_0 \nonumber \ .
\end{align}
Notice that in the massless limit the first, fourth and fifth equation of this system reproduce the analogous equations in 11d supergravity given by \cite[Eq.(9.71)--(9.73)]{gauntlett-macconamnha-mateos-waldram}. The third fixes the function $f$ in terms of the dilaton and the second is solved imposing the on shell constraints \eqref{rule3}.

It so happens that the Bianchi identities for the fluxes are automatically satisfied. So (\ref{ode5}) is the complete system we need to satisfy in order to find an AdS$_4$ solution. 

Moreover, given a solution of (\ref{ode5}), one can always find another rescaled solution for which the curvature and string coupling are both small, so that the supergravity approximation we are using in this paper is justified. This can be done by using the transformations \cite[Eq.(4.2)--(4.3)]{gaiotto-t-6d}; the first is $F_0 \to n F_0$, $\phi\to \phi - \log n$, which is a symmetry of (\ref{ode5}); the second has to be supplemented with transformation law for $f$ and $g$: 
\begin{equation}\label{eq:resc}
	(A,\, f,\, g,\, \phi,\, x,\, r) \to (A+ \Delta A,\, e^{\Delta A} f,\, e^{\Delta A} g,\, \phi-\Delta A,\, x,\, e^{\Delta A}r)\ .
\end{equation}

In the next section we will see that a certain three-dimensional submanifold of the space of parameters is invariant under (\ref{ode5}); on that submanifold the system is then reduced to a much more manageable system, whose solutions are the main focus of this paper. In section \ref{attractor} we will then go back to the general system (\ref{ode5}), and find more solutions to it, albeit only numerically.


\section{Natural compactifications} \label{natural}

In this section we will obtain solutions which generalize to the massive case the ${\cal N}=1$ compactifications reviewed in section \ref{cft6sec}.

\subsection{Reducing the ODE system} 
\label{sub:red}

In section \ref{sub:ode5} we obtained the system (\ref{ode5}) of ODEs, which is necessary and sufficient to find an AdS$_4$ solution within our Ansatz. We are now going to impose a certain constraint on (\ref{ode5}), which will simplify it quite a bit. 

Originally we found this simplification by noticing empirically that many solutions had a constant ratio between the functions $g$ and $e^A$ in (\ref{433metric}), which are the ``radii'' of $\Sigma_3$ and of AdS$_4$ respectively. A posteriori this assumption is quite natural, and indeed it was later found very useful for the AdS$_5$ solutions of \cite{afpt} as well (where it is called a ``compactification Ansatz''). A rough justification is as follows. The holographic dual of putting a CFT$_6$ on $\rr^3 \times \Sigma_3$ would consist in replacing $ds^2_{\rm AdS_7}= \frac{d \rho^2}{\rho^2} + \rho^2 ds^2_{\rr^6}$ with $\frac{d \rho^2}{\rho^2} + \rho^2 (ds^2_{\rr^3} + ds^2_{\Sigma_3})$. In the IR, if this leads to a CFT$_3$, one would expect that the $\rho^2$ in front of $ds^2_{\Sigma_3}$ somehow disappears; our Ansatz is somehow that it does not also get multiplied by a further function of $\Sigma_3$, or worse. 

So in practice we assume that $g e^{-A}$ is constant. As usual for a dynamical system, if one imposes a constraint one needs to worry about possible ``secondary constraints''; in our case, we need to check what happens when we use (\ref{ode5}) in $(g e^{-A})'=0$. We do get a secondary constraint: it turns out that $\frac{f e^{-A}}{\sqrt{1-x^2}}$ needs to be constant as well. In principle we could get now a third constraint as well, but imposing compatibility with (\ref{ode5}) of this second constraint we simply end up fixing both constants. This procedure actually only works when $\Sigma_3$ has Ricci scalar $R<0$; without loss of generality we then fix $R=-6$. The result is then
\begin{equation} \label{rule3}
f = \frac{2}{5} e^{A} \sqrt{1-x^2} \ , \qquad g = \frac{2}{\sqrt{5}}e^A \ . 
\end{equation}
In other words, within the five-dimensional space spanned by the parameters $(f,g,A,x,\phi)$, we have found a three-dimensional subspace that is left invariant by the flow. 
  
The system (\ref{ode5}) now simplifies quite a bit; after eliminating $f$ and $g$ using (\ref{rule3}), it only involves the warping factor $A$, the dilaton $\phi$ and the angle between the two six dimensional spinors $x=\cos \psi$. Moreover, two equations become redundant. The system then becomes
\begin{align} \label{ode34}
\phi^{'} & = \frac{1}{8} \frac{e^{-A}}{\sqrt{1-x^2}} \left( 21x - 6 x^3 + 2(5-2x^2) F_0e^{A + \phi}  \right) , \nonumber \\
x' & = \frac{1}{4} e^{-A}\sqrt{1-x^2} \left( 3x^2-8 + 2x F_0e^{A + \phi}  \right) \ ,  \\
A^{'} & = \frac{1}{8} \frac{e^{-A}}{\sqrt{1-x^2}} \left( 5x + 2e^{A + \phi} F_0 \right) \ . \nonumber
\end{align}
where the derivative with respect to the radial coordinate $r$ is denoted by $\partial_r (\  ) \equiv (\  )^{'}$. Notice that this system of ode's looks very similar to the corresponding BPS equations in AdS$_7$ given in \cite[Eq.(4.17)]{afrt}.
As we will see in section \ref{74sec}, this similarity can be made more explicit.

 
\subsection{Metric and fluxes} 
\label{sub:flux}

With the constraints (\ref{rule3}), the full ten-dimensional metric (\ref{433metric}) becomes
\begin{equation} \label{natural433met}
ds^2_{10}=e^{2A} \left( ds^2_{\rm AdS_4} + \frac{4}{5} ds^2_{\Sigma_3} \right)+dr^2 + \frac{4 (1-x^2) }{25} e^{2A} Ds^2_{S^2} \ .
\end{equation}
Notice the similarity with the AdS$_7$ metric (\ref{ads7m3}).

Let us also give the form of the fluxes here. Their general expression will be given in section \ref{attractor} below, but for the choice (\ref{rule3}) they are quite simple:
\begin{align} \label{fluxes3}
F_2 &= f e^{-\phi} (\omega_5 - \omega_1) -\frac{2F_0}{5} x f e^{A} \  \omega_1 \ , \nonumber \\
F_4 &= -\frac{2}{5} f e^{A-\phi}  (\omega_{23} + x f \ \omega_{15}) \ ,  \\
H &= - \frac{2}{5}e^A (dr\omega_5 + x f \ y^ie^i \omega_4) - \left( \frac{3(x^2-3)}{2} e^{-A} +x F_0 e^{\phi}  \right) {\rm vol}_{M_3} \ . \nonumber
\end{align}
Here, the forms $\omega_A$ were defined in (\ref{wi}). We left $f$ in these expressions, even though it should be thought of as given by (\ref{rule3}). These expressions are again very similar to the fluxes for the AdS$_7$ solutions of \cite{afrt}: there, $F_2$ only had a component along the volume of the $S^2$, which roughly corresponds to our $\omega_1$;  $H$ only had a component along the volume of the internal manifold $M_3$, which for us is ${\rm vol}_{M_3} = f^2 dr\omega_1$. In studying flux quantization for these fluxes, a crucial role will be played by the combination
\begin{equation}\label{eq:q}
 q \equiv f e^{-\phi} = {\rm radius} (S^2) \,e^{-\phi}\ ;
\end{equation}
using (\ref{rule3}) we see that it is very similar to the quantity of the same name in \cite[Eq.(4.41)]{afrt}.


\subsection{Supersymmetric maps} \label{74sec}

We have noticed already a few similarities with the AdS$_7$ solutions of \cite{afrt}. In particular, the ODE system (4.17) in that paper looks very similar to our (\ref{ode34}). Remarkably, the two systems are mapped into each other by:\footnote{The map might be more readable to some as
\begin{equation} 
e^{A_4} = \left(  \frac{5}{8} \right)^{3/4} e^{A_7} \ ,\qquad  e^{\phi_4} = \left(  \frac{5}{8} \right)^{1/4} \frac{e^{\phi_7}}{\sqrt{w}} \ , \qquad x_4 = \frac{x_7}{\sqrt{w}} \ ,\qquad  r_4 = \left(  \frac{5}{8} \right)^{1/4} r_7 \ ,
\end{equation}
with $w=\frac{5+3x_7^2}{8}$. In this paper we prefer to drop the indices $4,7$ that label the dimension of the internal space, just to make the expressions more readable.}
\begin{equation} \label{map}
\begin{split}
	e^{A} \rightarrow \left(  \frac{5}{8} \right)^{3/4} e^{A} \ ,\qquad  e^{\phi} \rightarrow \left(  \frac{5}{8} \right)^{1/4} \frac{e^{\phi}}{\sqrt{w}} \ , \\ 
	 x \rightarrow \frac{x}{\sqrt{w}} \ ,\qquad  r \rightarrow \left(  \frac{5}{8} \right)^{1/4} r \ ,\qquad F_0 \to - F_0 \ ,
\end{split}
\end{equation}
where we defined a warping function
\begin{equation} \label{w}
	w \equiv \frac{5+3x^2}{8}\ .
\end{equation}
Actually the requirement that (\ref{map}) should map (\ref{ode34}) in \cite[Eq.(4.17)]{afrt} leaves one parameter free, which we fixed by requiring that $q$ in (\ref{eq:q}), which reads $q_4 = \frac{2}{5} e^{A-\phi} \sqrt{1-x^2}$ transforms into the $q$ of \cite[Eq.(4.41)]{afrt}, $q_7 = \frac{1}{4} e^{A-\phi} \sqrt{1-x^2}$. As we anticipated, this will play a crucial role in the study of the flux quantization of section \ref{fluxquant}.

The map \eqref{map} acts on the full ten dimensional metric as
\begin{equation}\label{7to4metric}
\begin{split}
	e^{2A}ds^2_{\rm AdS_7}+ &dr^2 + \frac{1-x^2}{16} e^{2A} ds^2_{S^2}\ \to \ \\
	& \sqrt{\frac58} \left[ \frac58e^{2A} \left( ds^2_{\rm AdS_4} + \frac{4}{5} ds^2_{\Sigma_3} \right) + dr^2 + \frac{1-x^2}{2(5+3x^2)} e^{2A} Ds^2_{S^2} \right] \ ,	
\end{split} 
\end{equation}
One can indeed see that the massless metric (\ref{433spontaneous}) is of this form; however, \eqref{7to4metric} is now valid also for massive solutions. 

The map (\ref{map}) also inspired a similar map for the AdS$_7$ to AdS$_5$ compactifications on Riemann surfaces \cite[Sec.~5.2]{afpt}. Combining the two maps we get:\footnote{Again an alternative way of presenting the map is:
\begin{equation} 
e^{A_4} = \left(  \frac{5}{6} \right)^{3/4} e^{A_5} \ ,\qquad  e^{\phi_4} = \left(  \frac{5}{6} \right)^{1/4} \frac{e^{\phi_5}}{\sqrt{\tilde{w}}} \ , \qquad x_4 = \frac{x_5}{\sqrt{\tilde{w}}} \ ,\qquad  r_4 = \left(  \frac{5}{6} \right)^{1/4} r_5 \ ,
\end{equation}
with $\tilde{w}=\frac{1+5x_5^2}{6}$, where the indices $4,5$ label the dimension of the AdS factor.
}
\begin{equation} \label{45map}
\begin{split}
	e^{A} \rightarrow \left(  \frac{5}{6} \right)^{3/4} e^{A} \ ,\qquad  e^{\phi} \rightarrow \left(  \frac{5}{6} \right)^{1/4} \frac{e^{\phi}}{\sqrt{\tilde{w}}} \ , \\ 
	 x \rightarrow \frac{x}{\sqrt{\tilde{w}}} \ ,\qquad  r \rightarrow \left(  \frac{5}{6} \right)^{1/4} r \ ,\qquad F_0 \to - F_0 \ ,
\end{split}
\end{equation}
where we introduced a new warping function
\begin{equation} \label{wtilde}
\tilde{w} \equiv \frac{5+x^2}{6} \ .
\end{equation}
Again, the transformation law for the full metric is
\begin{equation}\label{5to4metric}
\begin{split}
	e^{2A} \left( ds^2_{\rm AdS_5} + ds^2_{\Sigma_2} \right)+ &dr^2 + \frac{1-x^2}{9} e^{2A} ds^2_{S^2}\ \to \ \\
	& \sqrt{\frac56} \left[ \frac56e^{2A} \left( ds^2_{\rm AdS_4} + \frac{4}{5} ds^2_{\Sigma_3} \right) + dr^2 + \frac{2(1-x^2)}{3(5+x^2)} e^{2A} Ds^2_{S^2} \right] \ ,	
\end{split} 
\end{equation}
Once again the map has one free parameter which is fixed by requiring that $q_4 = \frac{2}{5} e^{A-\phi} \sqrt{1-x^2} $ transforms into $q_5 = \frac{1}{3} e^{A-\phi} \sqrt{1-x^2}$. With this choice $q$ is a universal quantity for the AdS$_7$ solutions and all of their compactifications:
\begin{equation}\label{eq:qqq}
q_4 = q_5 = q_7 \ .
\end{equation}

In summary, the result of this section is that there is a one-to-one correspondence between solutions of the reduced BPS system (\ref{ode34}) and solutions of the BPS system for AdS$_7$ solutions in \cite[Eq.(4.17)]{afrt}. Moreover, \cite[Sec.~5.2]{afpt} establishes that there is a one-to-one correspondence of AdS$_7$ solutions with AdS$_5\times \Sigma_2$ solutions, with $\Sigma_2$ a Riemann surface: 
\begin{equation}\label{eq:111}
	{\rm AdS}_4\times \Sigma_3 \ \leftrightarrow \ {\rm AdS}_7 \ \leftrightarrow \ {\rm AdS}_5\times \Sigma_2\ .
\end{equation}
The correspondence to AdS$_5$ will be important for us, because in \cite{afpt} the BPS system was solved analytically, as we will review in section \ref{massivesolutions}. 

However, before we are able to claim that (\ref{eq:111}) is also a correspondence between solutions, we should also check that flux quantization is respected by it. We will do so now, and we will then look at concrete analytic solutions.

\subsection{Flux quantization} \label{fluxquant}

Flux quantization is formally very similar to the discussion in \cite[Sec.~4.8]{afrt} and in \cite[Sec.~5.4]{afpt}; here we will summarize the results of that discussion, and refer to those papers for details.

For massive solutions, as usual the Bianchi identity (away from sources) $dF_2 = H F_0$ implies that $H$ can be rewritten as $H=\frac{F_2}{F_0}$. As a consequence the $B$ field takes the form:
\begin{equation}
B = \frac{F_2}{F_0} + b \ ,
\end{equation} 
where $b$ is a closed two-form to be determined by imposing flux quantization. We can limit ourselves to considering $b$ of the form 
\begin{equation}
	b = b_0 (\omega_1 - \omega_5)\ ,
\end{equation} 
which is indeed closed, as shown in section \ref{twistedforms} (recall that we have normalized the Ricci scalar to $R=-6$). We can thus rewrite the $B$ field as
\begin{equation}\label{eq:Bext}
	B = \left(-\frac{q}{F_0} + b_0\right)(\omega_1 - \omega_5) - \frac{2}{5} q x e^{A+\phi} \omega_1 \ .
\end{equation}
One should also recall that $B$ is not a two-form; it can transform on intersections of open sets by ``large gauge transformations'', namely closed two-forms whose periods are integer multiples of $4\pi^2$.   

As for the RR fluxes, the Romans mass satisfies $F_0= \frac{n_0}{2\pi}$, $n_0 \in \mathbb{Z}$; also, the ``twisted'' fluxes
\begin{equation}
	\tilde F_2 \equiv F_2 - B F_0 \ ,\qquad \tilde F_4 \equiv F_4 - B \wedge F_2 + \frac12 B\wedge B F_0 \ 
\end{equation}
should have integer periods. The two-form is
\begin{equation}
	\tilde F_2 = - b F_0= - b_0 F_0 (\omega_1-\omega_5)\ ;
\end{equation}
flux quantization now implies 
\begin{equation}\label{eq:b0}
	b_0 = -\frac{n_2}{2 F_0}\ .
\end{equation}
The four-form $\tilde F_4$ can be written as
\begin{equation}\label{eq:F4t}
\tilde{F}_4 =  \frac{1}{F_0}\left( q^2 - \frac{n_2^2}{4} \right) \omega_{15} -\frac{1}{9} dy\ y^i e^i \omega_{2}  \ ,
\end{equation}
or also as $\tilde F_4 = d \tilde C_3$, where\footnote{It is interesting at this point to compare our fluxes to their AdS$_5$ counterpart in \cite{afpt}. For example, (\ref{eq:C3}) is formally the same as in \cite{afpt}; the form $y^i e^i \omega_2= y^i e^i e^j Dy^j$, if one now declares $i$ to be only $1,2$, becomes $e^1 e^2 (y^1 D y^2 - y^2 D y^1)= \sin^2(\theta) D \psi {\rm vol}_{\Sigma_2}$, which reproduces the expression in \cite[Sec.~5.3]{afpt}. As another example, in the expression for $F_2$ in (\ref{fluxes3}), $\omega_1$ is simply the covariantized ${\rm vol}_{S^2}$, and $\omega_5=\frac12 \epsilon^{ijk}y^i e^{jk}$ becomes $y^3 e^1 e^2= \cos(\theta){\rm vol}_{\Sigma_2}$, which reproduces the $F_2$ in \cite[Sec.~5.1]{afpt}.}
\begin{equation}\label{eq:C3}
	\tilde{C}_3 =\frac{1}{2 F_0} \left( q^2 - \frac{n_2^2}{4} \right) y^i e^i \omega_2 \ .
\end{equation}
We made use of the derivation rule $d( y^i e^i \omega_2 ) = 2 \omega_{15}$ which descends from equation \eqref{formsderivatives}, and we also inserted the relation $\frac{d(q^2)}{dy} = -\frac29 F_0 $, which can be verified using the BPS equations \eqref{ode34} together with the radial change of coordinates $dr = \frac{5}{18 q e^A} dy$; we will later find it again in (\ref{eq:odeb}). Near a regular point, regularity of $B$ and $F_2$ implies that $n_2$ should be zero, and that $q\to 0$. Moreover, one can see from (\ref{ode34}) that $q$ starts linearly in the radial coordinate, so that in the end $\tilde C_3 \sim r^2 y^i e^i e^j D y^j= x^i e^i e^j Dx^j$, where now the $x^i \equiv r y^i$ are coordinates on $\rr^3$; so $\tilde C_3$ is a regular form, and $\tilde F_4$ has no periods in this case. In presence of sources, the discussion changes a bit. Flux quantization now requires the flux integrals to be integer for cycles that do not intersect the sources. We can take such cycles to be at fixed $y$; then the only relevant term in (\ref{eq:F4t}) is $\omega_{15}= \frac12{\rm vol}_{S^2} \epsilon^{ijk}y^i e^{jk}$, whose integral vanishes because $\int_{S^2} y^i=0$.

We can now start introducing D8-branes, which we will allow to also have D6-charge; so, across such a brane both fluxes $(n_0,n_2)$ will jump to new values $(n_0',n_2')$. The ``slope'' $\mu\neq \frac{\Delta n_2}{\Delta n_0}= \frac{n_2'-n_2}{n_0'-n_0}$ is an integer. Imposing that (\ref{eq:Bext}) be continuous we find the condition
\begin{equation}\label{eq:qd8}
	[q]_{r=r_{D8}} = \frac{n_2' n_0 - n_2 n_0'}{2(n_0'-n_0)} =  \frac12(-n_2 + \mu n_0) = \frac12 (-n_2' + \mu n_0') \ ,
\end{equation}
which is formally identical to \cite[Eq.(4.45)]{afrt}. We now understand why we chose to have the map keep $q$ invariant across dimensions, (\ref{eq:qqq}). 

We finally have to understand what happens to flux quantization of $H$. This is complicated by the fact that in presence of D8's one might have a region of space where $F_0=0$, where (\ref{eq:Bext}) does not apply; one then has to use a separate expression for $B$ in the massless solution. A lengthy discussion \cite{afpt} (actually obtained by the present authors in collaboration with the authors of \cite{afpt}) establishes that 
\begin{equation}\label{eq:Hfluxq}
	N \equiv -\frac1{4\pi^2}\int H = (|\mu_n| + |\mu_{n+1}|) + \frac1{4\pi} e^{2A(x=0)}(|x_n|+|x_{n+1}|)\ ,
\end{equation}
refining an earlier analysis in \cite{gaiotto-t-6d}. Here the indices ${}_n$ and ${}_{n+1}$ refer to the D8 brane right before and right after the region where $F_0=0$. If that region does not exist, then $N=(|\mu_n| + |\mu_{n+1}|)$, as already remarked in \cite{gaiotto-t-6d}.

It can now be checked with some patience that the condition in (\ref{eq:Hfluxq}) is reproduced also in AdS$_4$, once one uses the map (\ref{map}). In \cite{afpt} this is also checked for AdS$_5$ solutions. 

In summary, we can conclude that the one-to-one correspondence (\ref{eq:111}) respects flux quantization. Thus it is a correspondence between string theory solutions, and not just supergravity solutions. 

In the following sections we will start studying concrete analytic solutions, reaping the rewards of the analysis performed so far.

\subsection{Massless solutions} \label{massless}

As a warm-up, we will first discuss the massless solution. This can be obtained directly as a solution of (\ref{ode34}), or applying the map \eqref{map} to the AdS$_7$ massless solution \ref{ads7massless10d}. Either way, one reproduces the ten-dimensional reduction (\ref{433spontaneous}) of the eleven-dimensional background found in \cite{gauntlett-macconamnha-mateos-waldram}, which in turn lifts the seven-dimensional gauged supegravity solution found in \cite{pernici-sezgin}. The metric reads
\begin{equation} \label{masslessmetric}
ds^2_{10}  = \left( \frac{5}{8} \right)^{\frac{3}{2}} \frac{R^3}{2} \sin \alpha \left[ ds^2_{\rm AdS_4} + \frac{4}{5} ds^2_{\Sigma_3} + \frac{2}{5} d\alpha^2 + \frac{4}{5} \frac{\sin^2 \alpha}{3 \cos^2 \alpha+5} Ds^2_{S^2}\right] \ .
\end{equation}
The same logic as in \cite[Sec.~5.1]{afrt} reveals that at the two poles $x \rightarrow \pm 1$ we have a D6 and a $\overline{\rm D6}$ stack. Indeed near the north pole $\alpha = 0$ the metric behaves as $ds^2_{M_3} \sim \alpha (d\alpha^2 + \frac{1}{4} \alpha^2 Ds^2_{S^2} )$ , which can be mapped into the usual metric describing a D6 in flat space $ds^2_{M_3} \sim \rho^{-\frac{1}{2}} (d\rho^2 +\rho^2 Ds^2_{S^2} )$ via the coordinate transformation $\rho = 2^{-\frac{4}{3}} \alpha^2$.  The dilaton is given by
\begin{equation}
e^{2\phi}  = \left(\frac{5}{8} \right)^{\frac{1}{2}} \frac{R^3 \sin^3\alpha}{5+3\cos^2\alpha} \ .
\end{equation}
For completeness we also give the expressions for the fluxes, which can be obtained applying the map (\ref{map}) to the solution (\ref{ads7massless10d}) and substituting the result into \eqref{fluxes3}:
\begin{equation}\label{fluxes30}
	F_2 = \frac{1}{2}(\omega_5 - \omega_1) \ ,\qquad
	F_4 = -\frac{R^3\sin \alpha}{32} d \alpha\, y^i e^i \omega_2 - \frac{R^3\cos \alpha \sin^2 \alpha}{8 (5+3\cos^2 \alpha)}  \omega_{15} \ . \\
\end{equation}
It is also possible to derive a simple expression for the $B$ field:
\begin{equation}
B =  \frac{R^3 \cos \alpha}{16}\ \omega_5 -  \frac{R^3 \cos \alpha (9-\cos^2 \alpha)}{16(5+3\cos^2\alpha)}  \omega_1 \ .
\end{equation}
We actually used this expression in checking that (\ref{eq:Hfluxq}) is also the correct flux quantization condition for AdS$_4$.

\subsection{Massive solutions} \label{massivesolutions}

We will now turn our attention to massive solutions. Thanks to the maps in section \ref{74sec} and to results obtained for AdS$_5$ solutions in \cite{afpt}, we will be able to provide many analytic solutions. In this section we focus our attention on massive solutions with D6 and O6 sources; solutions with D8-branes will be shown in section \ref{sub:D8}. 

In \cite{afpt} the BPS system of ODEs was actually solved \emph{analytically}. We can then use the map (\ref{45map}) to provide a solution to our BPS system (\ref{ode34}) as well. The solution is
\begin{equation}\label{eq:beta}
	e^A = \frac{5^{3/4}}6\left(-\frac{\del_y\beta}{2y}\right)^{1/4} \ ,\qquad 
	x= \sqrt{\frac{-2 y\del_y\beta}{5 \beta - 2 y \del_y\beta}} \ ,\qquad
	e^\phi= \left(\frac52\right)^{1/4}\frac{(-\del_y\beta/y)^{5/4}}{12 \sqrt{5 \beta - 2 y \del_y\beta}}\ ,
\end{equation} 
where $y$ is defined by $\frac{dr}{dy}= \left(\frac65\right)^2 \frac{e^{3A}}{\sqrt{\beta}}$, and $\beta$ is a solution of the equation
\begin{equation}\label{eq:odeb}
	\del_y (q^2)= \frac{F_0}{72}\ ,\qquad q=\frac{y \sqrt{\beta}}{\del_y \beta}\ ;
\end{equation}
the expression of $q$ is obtained from its definition (\ref{eq:q}) and (\ref{eq:beta}). 
This equation can be easily solved by $\frac{\beta}{(\del_y \beta)^2}= \frac1{72} F_0\frac{y-\hat y_0}{y^2}$. 

Before showing some examples, let us comment on the regularity of these solutions. For compactness, we need the $S^2$ in \eqref{natural433met} to shrink in two points, that we think of as a ``north pole'' and a ``south pole''. The way this can be done was analyzed in \cite{afrt} for the AdS$_5$ solutions; the results can be applied directly to our AdS$_4$ case as well. This can be read off from the map (\ref{5to4metric}); basically, the function $\frac{2(1-x^2)}{3(5+x^2)}$, multiplying the factor of $ds^2_{S^2}$ after applying the map, goes to the same factor $1/9$ as the factor $\frac{1-x^2}9$ before applying the map. So the leading behavior does not change, and we can copy the results in \cite[Sec.~5.3]{afpt}. 

The results can be summarized as follows. The local behavior around a pole is associated to features of the function $\beta$ in (\ref{eq:beta}):
\begin{itemize}
	\item A single zero of $\beta$ corresponds to a regular point.
	\item A double zero of $\beta$ corresponds to presence of a stack of D6-branes. 
	\item A square root behavior $\beta\sim \beta_0 + \beta_{1/2} \sqrt{y-y_0}+\ldots$ corresponds to presence of an O6-plane.
\end{itemize}

The O6 case perhaps requires a few more comments. First of all, in that case it is understood that one needs to mod out the solution by worldsheet parity $\Omega_{\rm ws}$ times a $\zz_2$ involution which acts on the $S^2$ as the antipodal map $\sigma$; the O6 will sit at what we called the pole, which is where the $S^2$ shrinks and $\sigma$ has a fixed point. To be more precise about what we mean by the presence of the O6-plane, it is perhaps instructive to look at the O6 in flat space, whose metric reads $H^{-1/2} ds^2_{\rr^6} + H^{1/2} (dr^2 +r^2 ds^2_{S^2})$, where $H=1-\frac{r_0}r$. One might be tempted to say that the O6 is located at $r=0$, but this would not make sense: the metric is in fact purely imaginary in the ``hole'' $r\le r_0$. The locus where the $S^2$ shrinks is in fact $r=r_0$. The square root behavior we described above is the same as around this $r=r_0$ point in flat space.

It is also worth pointing out that this is really the same behavior as for an O6 in flat space, and not the one found in \cite{saracco-t}. In that paper, the singularity of an O6 in presence of Romans mass was replaced by a wormhole-like behavior. There is no contradiction: the geometry of that case was very different, with the parallel directions fibred in a certain way over the transverse $S^2$. It was inspired by the solution in \cite{dewolfe-giryavets-kachru-taylor}, for which it meant to provide a local (but non-smeared) version.

Just like O6's, also D6-branes can only occur at the north or south pole.  For AdS$_7$ solutions, this could simply be explained by the SO(3) symmetry. In our case, one can explain this through calibrations. Just as in \cite{afrt}, the calibration for a D6 is the variable $x$. A D6 can only sit where $x=1$ (and an anti-D6 where $x=-1$). Imagine starting from a solution with a regular point; in that case it can be seen from (\ref{rule3}) that 
\begin{equation}\label{eq:x1}
	x\to \pm1
\end{equation}
at the poles. So a probe D6 will want to sit there. It can actually be checked with some more work that even for solutions where D6's and/or O6's are already present $x\to \pm1$ at the poles.

After these preliminary remarks, let us see some examples of analytic solutions. 

\subsubsection{Solution with one D6 stack}\label{ssub:1D6}

The first solution we will analyze has a regular south pole, and a stack of D6's at the north pole. This corresponds to a $\beta$ with a double zero and a single zero.

This solution is obtained by applying the map \eqref{45map} to the AdS$_5$ solution in \cite[Sec.~5.5]{afpt}. We can also obtain it by applying (\ref{eq:beta}) and (\ref{natural433met}) to
\begin{equation}\label{eq:betaD6}
	\beta=\frac8{F_0}(y-y_0)(y+2 y_0)^2 \ ,
\end{equation}
which is the simplest analytic solution of (\ref{eq:odeb}). Given our general analysis, we expect a regular point at $y_0$, and a D6 stack at $-2y_0$.

The metric reads
\begin{equation} \label{D6regmet}
ds^2_{10} = \sqrt{\frac{5(y+2y_0)}{3F_0}} \left[ \frac56 ds^2_{\rm AdS_4} + \frac23 ds^2_{\Sigma_3}
+ \frac14\frac{dy^2}{(y_0-y)(y+2y_0)} + \frac{2}{3} \frac{(y_0-y)(y+2y_0)}{y^2-5y_0y+10y_0^2} Ds^2_{S^2} \right]\ ,
\end{equation}
which is valid for $F_0 >0$ if the new radial coordinate has range $y \in [-2y_0,y_0]$. The dilaton is determined to be
\begin{equation}\label{eq:D6regphi}
 e^{4\phi} = \frac{15}{F_0^3} \frac{(y+2y_0)^3}{(y^2-5y y_0+10y_0^2)^2} \ .
\end{equation}

As a cross-check, let us analyze the local behavior of the metric around the poles, which are defined as the two end points of the interval where the $S^2$ shrinks: $\{y=y_0 , y=-2y_0 \}$. 
\begin{itemize}
\item Around $y_0$ the metric is proportional to $ds^2_{M_3} \sim \frac{dy^2}{4(y_0-y)} + (y_0-y) Ds^2_{S^2}$, which can be mapped into the flat space metric $d\rho^2 + \rho^2 Ds^2_{S^2}$ with a simple coordinate transformation $\rho = \sqrt{y_0-y}$. In other words the solution is regular around this pole.
\item  The D6 singularity is still present at the second pole $y=-2y_0$, where the local behavior of the metric is $ds^2_{M_3} \sim \frac{dy^2}{\sqrt{y+2y_0}} + (y+2y_0)^{\frac{3}{2}} Ds^2_{S^2}$. Indeed if we define a new radial coordinate $\rho = y+ 2y_0$ we recover the metric describing the neighborhood of a D6 in flat space: $ds^2_{M_3} \sim \rho^{-\frac{1}{2}} (d\rho^2 +\rho^2 Ds^2_{S^2} )$. 
\end{itemize}

For completeness it is also worth giving the full expressions for the fluxes:
\begin{align}\label{eq:flux-D6}
F_2 & = q (\omega_5 - \omega_1) + \frac{2 q y (y+2y_0)}{y^2-5 y y_0 + 10y_0^2 } \omega_1  \ , \nonumber \\
F_4 & = -\frac{1}{9} dy \, y^i e^i \omega_2  + \frac{2 q y (y+2y_0)}{F_0 (y^2-5 y y_0 + 10y_0^2) } \omega_{15} \ , \\
H & = -\frac{1}{9 q} dy \omega_5 +  \frac{2 q y (y+2y_0)}{F_0 (y^2-5 y y_0 + 10y_0^2)} y^ie^i \omega_4
-\frac{q(y+2y_0)(19y^2+65yy_0-90y_0^2)}{2F_0 (y^2-5 y y_0 + 10y_0^2)^2} dy \omega_1 \ , \nonumber
\end{align}
which are expressed in terms of the function $q = \frac{1}{3} \sqrt{2F_0 (y_0-y)}$.

Flux quantization can be analyzed by direct inspection of (\ref{eq:flux-D6}), or by applying our general results in section \ref{fluxquant}. The result is that the parameter $y_0$ is fixed to be 
\begin{equation}\label{eq:y0n2}
	y_0=\frac38 \frac{n_2^2}{F_0}\ ;
\end{equation}
$n_2$ is the number of D6 in our stack at the north pole $y=-2y_0$. Replacing this expression in (\ref{D6regmet}) and defining
\begin{equation}
	\tilde y = \frac y{y_0}
\end{equation}
one recovers the metric (\ref{eq:intro-D6}). However, (\ref{D6regmet}) also appears as a piece of metrics with D8's, and in that case $y_0$ is fixed to a different value, as we shall see.

Meanwhile, from (\ref{eq:y0n2}), (\ref{D6regmet}), (\ref{eq:D6regphi}), we can also check explicitly that by taking $n_2$ to be large we can make both the curvature and the string coupling small. This is in agreement with the general observation made around (\ref{eq:resc}); notice that that transformation preserves the constraint (\ref{rule3}).

\subsubsection{General massive solution}\label{ssub:general}

We will now show a more general solution; without D8-branes, this is in fact the most general one. It can be obtained by applying the map \eqref{45map} to the AdS$_5$ solution in \cite[Sec.~5.6]{afpt}, or by applying (\ref{eq:beta}) and (\ref{natural433met}) to
\begin{equation}\label{eq:gen}
	\beta = \frac{y_0^3}{b_2^3 F_0}\left(\sqrt{\hat y} - 6\right)^2\left(\hat y + 6 \sqrt{\hat y}+6b_2 - 72 \right)^2 \ , 
\end{equation}
where
\begin{equation}
	\hat y \equiv 2 b_2 \left(\frac y{y_0} -1\right)+36\ . 
\end{equation}
The parameter $b_2$ has the interpretation of $b_2 \equiv \frac{F_0}{y_0} \beta_2$, where $\beta_2$ is half the second derivative of $\beta$ in $y_0$. The resulting solution is 
\begin{subequations}
	\begin{equation}
	e^{8\phi} =  \frac{ \left(\frac56\right)^2 b_2^{11} \beta^3}{\hat y^3 F_0^3 y_0^{11}(4(b_2 -18)^2+30(b_2-12)\sqrt {\hat y} + (b_2-18)\hat y + \hat y^2)^4 }\ ,
	\end{equation}
	\begin{equation}
		ds^2_{M_3}= \sqrt{\frac56}\left(\frac{y_0^5}{ 2^8 b_2^5 F_0^3 \hat y^3 \beta}\right)^{\frac14}d\hat y^2+ 
		\frac{ \frac{1}{9}\sqrt{\frac56} \left(\frac{ b_2^7 \beta^3 F_0 \hat y}{y_0^7}\right)^{\frac14} Ds^2_{S^2}}
		{4(b_2 -18)^2+30(b_2-12)\sqrt {\hat y} + (b_2-18)\hat y + \hat y^2}\ ,
	\end{equation}	
\end{subequations}
The meaning of this solution depends on the parameter $b_2$. Summarizing the analysis in \cite[Sec.~5.6]{afrt}:
\begin{itemize}
	\item If $b_2<12$, $\beta$ has two double zeros, so the solution corresponds to two D6 stacks, one at $\hat y = \sqrt{-3+\sqrt{81-6 b_2}}$, one at $\hat y = 36$. 
	\item If $b_2 >12$, the solution corresponds to a D6 stack at one pole $\hat y=0$ and an O6 singularity at $\hat y= 36$.  
	\item If $b_2 = 12$, $\beta$ simplifies to $\frac{y_0^3}{1728 F_0} \hat y (\hat y-36)^2$, which is (\ref{eq:betaD6}) up to coordinate change; so this case corresponds to a single D6 stack at $\hat y=36$. 
\end{itemize}
All this agrees with the qualitative analysis performed in \cite{afrt} for the AdS$_7$ solutions.

\subsection{Massive solutions with D8's}\label{sub:D8}

One can also obtain metrics with arbitrary numbers of D8-branes. These solutions are a bit more subtle: the Romans mass $F_0$ will jump across the D8-branes are located, and as a result the expression of the metric will change. Despite the jump in the Romans mass, the full metric can be made continuous by tuning the parameters properly. In other words we have to piece together solutions we have already studied. The position of the D8's is then fixed by (\ref{eq:qd8}). 

In \cite[Sec.~5.7]{afpt}, the procedure was illustrated with two examples, with one and with two D8-branes. It is now easy to apply the map (\ref{45map}) to those solutions. 

The solution with one D8 consists of two copies of (\ref{D6regmet}), glued exactly as in \cite[Eq.(4.42),(4.45)]{afpt}. We will not repeat it here.

Here is instead a solution with two D8's, which is the AdS$_4$ compactification of the AdS$_7$ solution obtained numerically in \cite[Fig.~5]{afrt}; the analytic expression for the AdS$_7$ solution is also given in \cite[Sec.~5.7]{afpt}. The configuration is symmetric, in the sense that the flux integers before the first D8 stack are $(-n_0<0,0)$, between the two stacks $(0,n_2=-k<0)$, and after the second stack $(n_0,0)$. We will assume $y_0<0$; the positions of the two D8 stacks will be $y_{\rm D8}<0$ and $y_{\rm D8'}=-y_{\rm D8}>0$. We get:
\begin{equation}\label{eq:2D8}
	\hspace{-.5cm}ds^2_{M_3}= \left\{ \begin{aligned} 
	\displaystyle 
		\sqrt{\frac{5(y+2y_0)}{48F_0}} \left(\frac{dy^2}{(y_0-y)(2y_0+y)}+ \frac83 \frac{(y_0-y)(2y_0+y)}{y^2-5y y_0+10y_0^2} Ds^2_{S^2}\right)
		 \ ,\qquad y_0<y<y_{\rm D8}\ ;\\
		\displaystyle 
		\frac{\sqrt{10(9^2R^6-32^2y^2)}}{32} \left( \frac{32^2 dy^2}{9^2 R^6 - 32^2 y^2} + \frac{2(9^2 R^6 - 32^2 y^2)}{5(9^2 R^6) + 3(32^2 y^2)} Ds^2_{S_2} \right)\displaystyle 
		\ ,\qquad y_{\rm D8}<y<-y_{\rm D8} ;\\
		\displaystyle 
		\sqrt{\frac{5(y+2y_0)}{-48 F_0}}\left(\frac{dy^2}{(y_0-y)(2y_0+y)}+ \frac83 \frac{(y_0-y)(2y_0+y)}{y^2-5y y_0+10y_0^2} Ds^2_{S^2}\right)
		\ ,\qquad -y_{\rm D8}<y<-y_0\ ;\\
	\end{aligned}\right.
\end{equation}
the metric in the middle region is the known massless metric in \ref{masslessmetric} after the change of coordinate $\cos\alpha = \frac{32}{9 R^3}y$. The parameter $R$, $y_0$, $y_{\rm D8}$ are also given in \cite[Sec.~5.7]{afpt}:
\begin{equation}
\begin{split}
	&R^6 = \frac{64}3 k^2\pi^2 (3 N^2 -4\mu^2) \ ,\\
	&y_0 = -\frac94 k \pi (N-\mu) \ ,\qquad
	y_{\rm D8} = -\frac94 k \pi (N-2 \mu)\ ,
\end{split}	
\end{equation}
where $\mu = \frac k{n_0}$.

As we mentioned, it is possible to generalize this solution to include an arbitrary number of D8-branes. It is also possible to include D6's or an O6 at the north and south pole, thus mixing the features of this section and of section \ref{ssub:general}.

\subsection{Summary and field theory interpretation} 
\label{sub:sum}

Let us summarize the solutions in this section, and make a few comments about their field theory interpretation. 

We have found an infinite class of $AdS_4\times M_6$ solutions, where $M_6$ is a fibration of $M_3$ over $\Sigma_3$; $M_3$ is topologically $\cong S^3$, while $\Sigma_3$ is a compact quotient of hyperbolic space. These solutions are in one-to-one correspondence (\ref{map}) with the AdS$_7$ solutions of \cite{afrt}. In particular, the metric on our $M_3$ is related to the internal manifolds in those AdS$_7$ solutions in the simple way (\ref{7to4metric}). It is a fibration of a round $S^2$ over an interval, and as such it has SO(3) isometry group. 

Our main aim in this paper was to find AdS$_4$ solutions dual to twisted compactifications of the $(1,0)$ CFT$_6$ dual to the AdS$_7$ solutions. Because of the fibration structure of our solutions (which was part of our Ansatz), and of the one-to-one correspondence (which came out as a result), the solutions we found seem to be exactly what we were looking for. 

We can contrast once again our solutions with the known massless ones \cite{gauntlett-macconamnha-mateos-waldram}, this time from a field theory perspective. For the ${\cal N}=2$ solution (\ref{s4n2def}), the internal space has SO(2)$\times$SO(3) symmetry; twisting mixes the SO(3) factor with the SU(2) of local Lorentz transformations on $\Sigma_3$, and we are left with only the SO(2) factor, which is in fact the R-symmetry of the resulting ${\cal N}=2$ theory. For the ${\cal N}=1$ massless solution (\ref{s4n1def}), the internal space has SO(4)$={\rm SU}(2)_{\rm L}\times {\rm SU}(2)_{\rm R}$ symmetry, and twisting mixes the ${\rm SU}(2)_{\rm L}$ factor with the SU(2) of $\Sigma_3$, leaving an SU(2) which is a flavor symmetry. There is no R-symmetry because the CFT$_3$ is only ${\cal N}=1$ supersymmetric.

For our solutions (and indeed for the ten-dimensional reduction of the massless solution, studied in section \ref{sub:IIA}), the isometry of the internal space is already just SU(2); twisting mixes it with the SU(2) of $\Sigma_3$, so that in the end we have no flavor or R-symmetry. (Again this is in no contradiction with the fact that the CFT$_3$ is only ${\cal N}=1$.) From the point of view of the gravity solution, the metric (\ref{natural433met}) has an $S^2$ factor, but the fact that it is non-trivially fibred means that the total space does not have SO(3) isometries: the presence of the connection breaks it. Even looking at the fluxes (\ref{fluxes3}), we see that they contain the forms (\ref{wi}), which break the SO(3) of the $S^2$. We did find the (\ref{wi}) by defining the ``twisted symmetry'' (\ref{eq:SO3D}), but that cannot be considered an isometry: it is a mix of a local Lorentz transformation (which happens point by point on $\Sigma_3$) and of an internal rotation.

Let us also point out however one point about the number of degrees of freedom of the CFT$_3$, which parallels a similar observation in \cite{afpt}. One can count the number of degrees of freedom of a CFT$_d$ via the coefficient ${\cal F}_{0,d}$ in the free energy ${\cal F}_d = {\cal F}_{0,d} T^d {\rm Vol}$, where $T$ is the temperature. Holographically this evaluates to the integral of $e^{5A-2 \phi}$ over $M_3$ for the CFT$_6$, and over $M_6$ for the CFT$_3$. Using the map (\ref{map}), one finds easily that 
\begin{equation}
	{\cal F}_{0,3}= \left(\frac 58\right)^4 {\cal F}_{0,6} {\rm Vol}(\Sigma_3)\ .
\end{equation}
In other words, the ratio of degrees of freedom is universal. Since the AdS$_7$ solutions are now analytic, one can evaluate ${\cal F}_{0,6}$ explicitly; this is indeed done in \cite{afpt} for an example. This might help find the CFT$_3$.

However, the CFT$_3$'s are only ${\cal N}=1$ supersymmetric, and have no flavor symmetry. For this reason, perhaps our solutions are more interesting as gravity solutions with localized sources; this was indeed our initial motivation. With this in mind, we will now return to our original system (\ref{ode5}), and see if we can find more interesting solutions, irrespectively of their field theory interpretation.


\section{Attractor solutions} \label{attractor}

In the last section we obtained a very large set of analytic solutions, in one-to-one correspondence with the AdS$_7$ solutions of \cite{afrt} and the AdS$_5$ solutions of \cite{afpt}; for this reason we called them ``natural compactifications''. The symmetric space $\Sigma_3$ needs to hyperbolic. 

In this section we will present another set of solutions, which depend on a larger number of parameters; we call them ``attractor solutions'', for reasons that will become clear. They are only known numerically. They exist for all values (positive, null, negative) of the curvature of $\Sigma_3$, although a positive sign appears to be preferred. 

The first sign that this class will be larger is that we will not impose the constraints \eqref{rule3} any longer. So we will have to revert to the system of five ODEs given in (\ref{ode5}). It will also not be possible any more to simplify the form of the metric like we did in \eqref{natural433met}, and we will have to keep the original form
\begin{equation} 
ds^2_{10}=e^{2A}ds^2_{\rm AdS_4} +g^2 ds^2_{\Sigma_3}  + dr^2 + f^2 Dy^i Dy^i.
\end{equation}

\subsection{Fluxes}

The fluxes will also not be given by (\ref{fluxes3}) any more. The general expression is instead
\begin{equation} \label{fluxes5}
\begin{split}
	F_2 &= ( -q + p F_0  )\ \omega_1 + (s + \tilde s F_0)\ \omega_5 \ ,  \\
	F_4 &= u\ \omega_{23} + (v + \tilde v F_0)\ \omega_{15} \ ,
\end{split}
\end{equation}
There are no new components in the fluxes with respect to the case (\ref{fluxes3}), but two terms acquire an additive contribution proportional to $F_0$. These extra contributions are proportional to each other:
\begin{equation}
\tilde v = (f^2 \cot \psi) \tilde  s \ , 
\end{equation}
and of course they vanish once we impose the constraint \eqref{rule3} (and fix $R = -6$). 
The other coefficients are as follows:
\begin{align}
& p = \frac{f^2 \left( -12 f g^2 e^{-A} -\sin \psi (R f^2 + 6 g^2 (1+\cos^2 \psi)) \right)}{\cos \psi (R f^2 -6g^2 \sin^2 \psi)}  \ ,\qquad u = f e^{-\phi} g^2 \left( \frac{\sin \psi}{f} - 3 e^{-A} \right)  \ , \nonumber \\
&  v = \frac{f e^{-\phi} (12 f g^2 e^{-A} - \sin \psi (R f^2 + 6 g^2)) }{6 \cos \psi} \ ,\ \tilde v = - \frac{f^2 g^2 (-12 f g^2 e^{-A} \sin \psi + R f^2 + 6 g^2 \sin^2 \psi )}{(R f^2 -6g^2 \sin^2 \psi)} \ .
\end{align}
Finally, $q= f e^{-\phi}$ and $s = -\frac{R}{6} q$, as in (\ref{eq:q}) and (\ref{fluxes3}).

Even though the fluxes are more complicated, flux quantization works similarly as in section \ref{fluxquant}. It is still true that $\tilde F_2 = -b F_0$, where now $b=-\frac{n_2}{2F_0}(\omega_1+\frac R6 \omega_5)$. 

The four-form flux quantization is also similar to section \ref{fluxquant}. We have $\tilde F_4=u\omega_{23}+ \tilde f_{15}\omega_{15}$, where $\tilde f_{15}=v + \tilde v F_0 +(-q + p F_0)(s+\tilde s F_0)+\frac{n_2^2 R}{24F_0}$. 
Closure implies $\tilde f_{15}' = 2 u$, so that we can still write $\tilde F_4 = d \tilde C_3$, where now $\tilde C_3 = \frac12 \tilde f_{15} y^i e^i \omega_2$. The coefficient $\tilde f_{15}$ reduces to the $\frac1{F_0}\left(q^2 - \frac{n_2^2}4\right)$ of (\ref{eq:C3}) when one imposes the constraint (\ref{rule3}) and fixes $R=-6$, but in general it is much more complicated. Nevertheless, upon substituting the local solution for a regular point in the next section, it still turns out that it starts quadratically in $r$, and that $\tilde C_3$ is regular around it. The rest of the flux quantization argument in section \ref{fluxquant} also runs in a similar way.

Finally, we are not going to consider solutions with a massless region (where $F_0=0$), so we will not need to work out the analogue of (\ref{eq:Hfluxq}) for $\int H$; we can simply use the formula $B= \frac{F_2}{F_0}+ b$. For cases without D8's, we can simply compute $B$ at the north and south pole and use Stokes' theorem. For solutions with D8's, one can use a logic similar to the solution with one D8 in \cite[Sec.~5.3]{afrt}, or \cite[Sec.~4.2]{gaiotto-t-6d}.

\subsection{Local Solutions: D6, O6, regular}

We will now assume that the internal space $M_3$ must have the topology of a $S^3$, namely the $S^2$ must shrink at the two extrema of the interval $[r_N,r_S]$, corresponding to the north and south pole. The shrinking of the $S^2$ implies that the function $f$ should vanish.

In the case of natural compactifications of section \ref{natural}, we discussed around (\ref{eq:x1}) that $x$ takes the values $\pm 1$ at the south and north pole. We will assume this to remain true for the present more general case as well. Since $x=\cos \psi$, this means that $\psi$ goes to $0$ and $\pi$ at the poles.

To complete the boundary conditions we have to specify how the function $f$ should go to zero at the pole, and this is what distinguish between the three different types of solution we are interested in.
\begin{enumerate}
	\item For a regular point, $f \sim r + O(r)^2$.
	\item Near D6, the metric $ds^2_{M_3}\sim \rho^{-1/2}(d \rho^2 + \rho^2 ds^2_{S^2})$; taking $r=\frac43 \rho^{3/4}$ gives $ds^2_{M_3}\sim dr^2 + \left(\frac	34 r\right)^2 ds^2_{S^2}$. So $f\sim \frac34 r$. 
	\item Near an O6, a similar computation gives $f\sim r^{1/5}+ O(r)^{2/5}$.
\end{enumerate}
We will now study more precisely these three cases. 

\subsubsection*{Regular Point}
We want to study the system around the boundary conditions at the north pole corresponding to a regular point: $[\psi=0, f=r]$. Thanks to translational invariance in $r$ we can assume $r_N = 0$ without any loss of generality and expand the functions entering the system \eqref{ode5} in power series. We determined the local solution up to order $r^3$. The precise expression was crucial as a boundary condition for the numerical analysis, but it is not very enlightening; we just summarize it as
\begin{equation}\label{eq:regexp}
\begin{split}
	& e^A  = e^{A_0} + O(r^2) \ ,\qquad e^{\phi}  =e^{\phi_0} +  O(r^2) \ ,\qquad g = g_0 + O(r^2) \ ,  \\
	& f = r + O(r^3) \ ,\qquad \cos \psi = 1 + O(r^2) \ .
\end{split}
\end{equation}
Notice that this expansion only involves odd and even functions in $r$. The parameter $\phi_0$ is fixed in terms of the other two parameters $g_0$ and $A_0$:
\begin{equation}\label{eq:regphi0}
	e^{\phi_0} = \frac{2 g_0 - \sqrt{6} \sqrt{84 g_0^2 -5 R e^{2A_0} } }{10 g_0 e^{A_0} F_0 } \ .
\end{equation}
So this is consistent if and only if $F_0 < 0$. In total, we have two free parameters in this boundary condition. One can check that $q$ and $p$ both $\to 0$.

\subsubsection*{D6 singularity}
We now switch to the boundary condition which is appropriate to describe a $D6$ singularity: $\psi\to 0$, $f\sim 3/4 r$. The leading behavior for the other fields can be inferred from the flat space D6, but it was not entirely clear how to continue the expansion; we determined it by trial and error, by imposing that the ODEs (\ref{ode5}) should be satisfied. We ended up with an expression where the leading behavior of each field is multiplied by an analytic function of $r^{4/3}$; for example, $f=\frac34 r \sum_k f_k r^{4k/3}$. We went up to $k=3$, obtaining again explicit expressions that would not tell the reader much. So as a summary let us just write
\begin{equation}\label{eq:D6exp}
\begin{split}
	& e^A  = r^{1/3} e^{A_0}+O(r^{5/3})  \ ,\qquad
	 e^{\phi}  = r e^{\phi_0} +O(r^{7/3})  \ ,\qquad
	 g  = r^{1/3} g_0+O(r^{5/3})\ ,  \\
	& f = \frac{3}{4} r + O(r^{7/3}) \ ,\qquad
	 \cos \psi  = 1 +O(r^{4/3}) \ .
\end{split}	
\end{equation}
So there are three free parameters in this boundary condition. For flux quantization, it is useful to compute that $q\to \frac34 e^{-\phi_0}$, $p \to 0$.

\subsubsection*{O6 singularity}
We finally consider an O6 singularity: $\psi\to 0$, $f \sim r^{1/5}$. Again by trial and error, we found this time a power series in $r^{4/5}$; for example, $f=r \sum_k f_k r^{4k/5}$. We went up to $k=2$. As a summary:
\begin{align}\label{eq:O6exp}
	& e^A  = r^{-1/5} e^{A_0} +O(r^{3/5}) \ ,\qquad
	 e^{\phi}  = r^{-3/5} e^{\phi_0} +O(r^{1/5}) \ ,\qquad
	g  = r^{-1/5} g_0 + O(r^{3/5}) \ ,\nonumber \\
	& f =r^{1/5} f_0 + O(r) \ ,\qquad
	 \cos \psi  = 1 + O(r^{4/5})  \ ,	
\end{align}	
where the parameter $\phi_0$ is fixed to be
\begin{equation}\label{eq:phi0O6}
	e^{\phi_0} = - (2 f_0) \frac{6 g_0 + \sqrt{36 g_0^2 -6 R e^{2A_0} } }{15 g_0 e^{A_0} F_0 } \ .
\end{equation}
This is again consistent if and only if $F_0 < 0$. In total, we have three parameters in this boundary condition too. For flux quantization purposes, it is useful to compute that $q\to 0$, $p \to \frac45 \frac{f_0^2}{F_0}e^{-\phi_0}$.

\subsection{Complete solutions}

We studied numerically the system (\ref{ode5}) with all three boundary conditions we discussed in the previous section, allowing the manifold $\Sigma_3$ to have positive, null and negative curvature. In what follows we present the possible solutions corresponding to $\Sigma_3 = S^3$, but the behavior is essentially the same for $T^3$ and $H_3$. We expected to have to perform some fine-tuning in order to obtain a physical solution, arriving at one of the same three boundary conditions at the other pole. Indeed one often ends up at the other pole with a singularity that we cannot interpret physically, where numerically one sees $f\sim r^{1/3}$, $g\sim r^{-1/3}$, $e^A\sim r^{-1/3}$. 

\begin{figure}[ht]
\centering	
	\subfigure[\label{fig:reg-reg}]{\includegraphics[scale=.8]{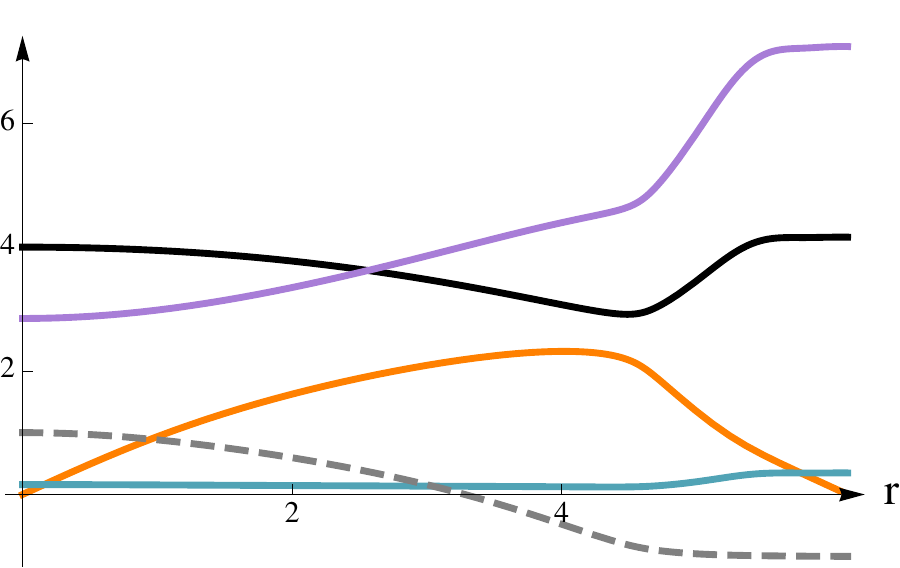}}
	\hspace{.4cm}
	\subfigure[\label{fig:D6-reg}]{\includegraphics[scale=.8]{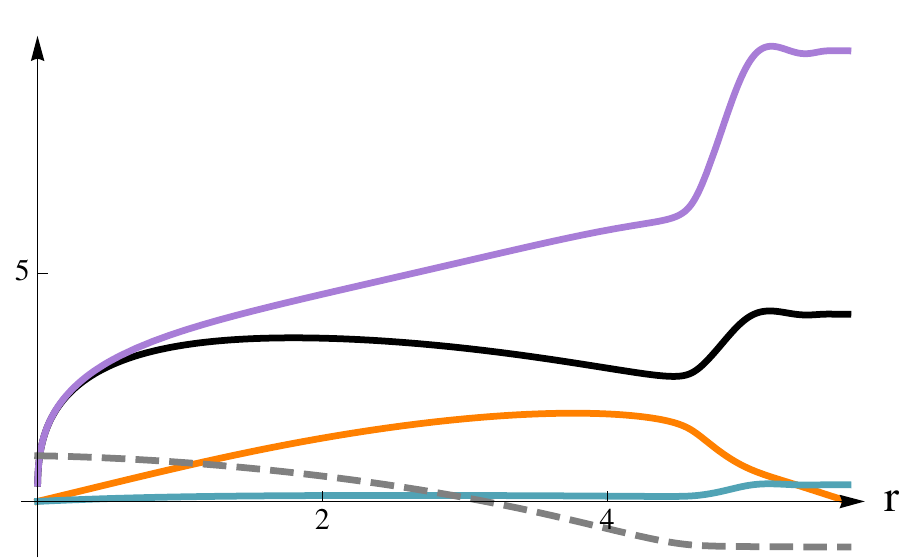}}
	\caption{Massive attractor solutions. In \subref{fig:reg-reg} we see a solution with two regular poles, and $n_0 = -10$  (as usual, $F_0=\frac{n_0}{2\pi}$). We plot $f$ (orange), $e^\phi$ (green), $e^A$ (black), $g$ (purple), $x=\cos \psi$ (dashed). In  \subref{fig:D6-reg} a solution with a stack of $n_2=10$ D6-branes at the north pole (left), and a regular point at the south pole (right); again $n_0=-10$, and $N=-\frac1{4\pi^2}\int H=-1$. In both cases, $R=6$, so $\Sigma_3=S^3$.}
	\label{fig:gen}
\end{figure}

\begin{figure}[ht]
\centering	
	\subfigure[\label{fig:O6-reg}]{\includegraphics[scale=.8]{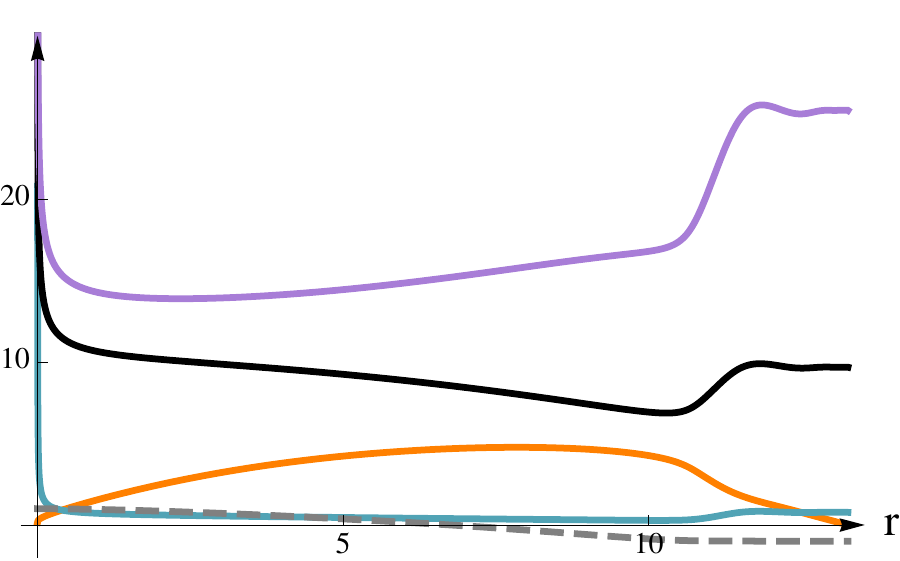}}
	\hspace{.4cm}
	\subfigure[\label{fig:D8}]{\includegraphics[scale=.8]{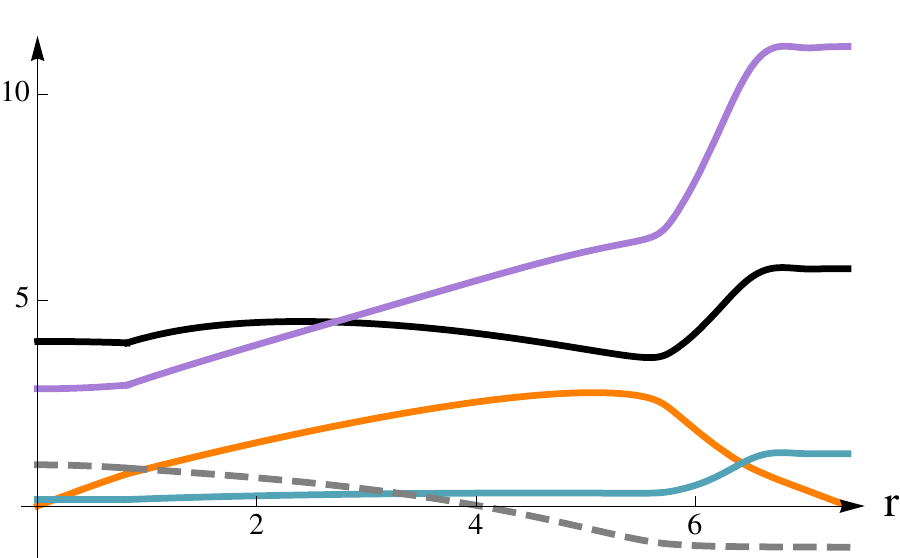}}
	\caption{Massive attractor solutions. In \subref{fig:O6-reg} we see a solution with an O6 at the north pole (left), and a regular point at the south pole (right).  In  \subref{fig:D8} a solution with two regular poles with a D8 stack in the middle (which is the sharp kink towards $r\sim 1$, most visible in the black and purple lines). In both cases, $R=6$, so $\Sigma_3=S^3$.}
	\label{fig:O6D8}
\end{figure}

Even more often, however, one in fact ends up more or less automatically at the other pole with a regular point. This happens for a large \emph{open set} in the space of the free parameters allowed by the boundary conditions of the previous section (two for the regular boundary condition, three for the D6 and O6). In most other cases, one has instead to perform a number of fine tunings. In the present case, the regular point appears to be an attractor. We show some examples of numerical solutions in figures \ref{fig:gen} and \ref{fig:O6-reg}. In all these cases, we started from the left with the relevant perturbative solution (schematically expressed in (\ref{eq:regexp}), (\ref{eq:D6exp}) and (\ref{eq:O6exp})), and continued numerically. The solution then ends by itself in a point where $f=0$ and the other functions go to constant values, which one can check to be consistent with (\ref{eq:regexp}), (\ref{eq:regphi0}) --- with a minimal modification due to $x$ being $-1$ rather than 1. Some solutions appear to display one or more mild kinks on the way to the attractor; one might worry about their effect on the curvature, but recall how (\ref{eq:resc}) can be used to make the curvature as small as one wishes.

It also appears to be equally easy to obtain solutions with D8-branes. Their position is again fixed by (\ref{eq:qd8}), and the attractor mechanism appears again at the south pole. An example is given in \ref{fig:D8}.

The O6 case in particular would appear promising to obtain examples with ``separation of scales''. In AdS$_4$ compactifications, the Kaluza--Klein scale $m_{\rm KK}$ is usually of the same order of the cosmological constant $\Lambda$, which is obviously unphysical. One might object that the negative sign of $\Lambda$ is even more unphysical. However, sometimes one manages to modify the AdS vacuum by adding some extra ingredient, which turns the cosmological constant positive \cite{kklt}; the lack of separation of scales might then be inherited by the resulting de Sitter as well\footnote{We thank T.~Van Riet for interesting discussions on this point.}. The presence of this phenomenon would also be interesting from the point of view of the CFT dual, since it would imply the presence of a large gap in operator dimensions. A few examples have been put forward where the is separation of scales (see for example \cite{dewolfe-giryavets-kachru-taylor,caviezel-etal,petrini-solard-vanriet}), but they usually rely on the smeared O6 we mentioned in the introduction (although see \cite{maxfield-mcorist-robbins-sethi} and the strategy in \cite{polchinski-silverstein}). With the simplest solution of figure \ref{fig:O6-reg}, which only has a single O6, we have not been able to achieve separation of scales, but by combining it with the other ingredients (D8-branes, and perhaps D6-branes at the other end) it might be possible. It would be interesting to explore this further.

\section*{Acknowledgments}
We would like to thank C.~Bachas, I.~Bena, N.~Bobev, J.~Maldacena, T.~Van Riet, A.~Zaffaroni for interesting discussions. We are also grateful to F.~Apruzzi, M.~Fazzi and A.~Passias, with whom we discussed several details of \cite{afpt}.  We are supported in part by INFN. A.~T. is also supported by the MIUR-FIRB grant RBFR10QS5J ``String Theory and Fundamental Interactions''  and by the European Research Council under the European Union's Seventh Framework Program (FP/2007-2013) -- ERC Grant Agreement n. 307286 (XD-STRING). 

\appendix

\section{$S^3$ to $S^2$ left-invariant forms projection} \label{s3s2}
In this appendix we introduce a formalism which will be useful to reduce the eleven-dimensional massless solution described in \cite[Sec.~9.4]{gauntlett-macconamnha-mateos-waldram}. In particular, we want to rewrite the $S^3$ metric in Hopf coordinates and reduce it to $S^2$. Our starting point is the metric: $ds^2_{3} =\frac{1}{4} \mu^i \mu^i$,
describing an $S^3$ fibered over a three manifold $\Sigma_3$. The triplet one-forms $\mu^i$ are defined as: $\mu^i = \sigma^i - \omega^i$,
with $\sigma^i$ being the left invariant forms on $S^3$ satisfying $d\sigma^i = \frac{1}{2}\epsilon^{ijk} \sigma^{jk}$, and $\omega^i$ is the connection of the fiber bundle $S^3 \rightarrow \Sigma_3$, related to the spin connection of the base space by $\omega^i = \frac{1}{2}\epsilon^{ijk} \omega^{jk}$.

Our goal is to compute the components of the one-forms $\mu^i$ along the $S^2$. This is achieved by introducing a parallel and orthogonal projectors:
\begin{equation}
P_{\parallel}^{ij} = \delta^{ij} - y^i y^j \ , \qquad P_{\perp}^{ij} = y^i y^j \ ,
\end{equation}
which satisfy $P_{\parallel} + P_{\perp} = 1$, where $y^i$ are the spherical harmonics that parametrize the $S^2$, (\ref{eq:y}). The corresponding decomposition for the one-forms $\mu^i$ is the following:
\begin{equation} \label{mudec}
\mu^i= \epsilon^{ijk}y^j Dy^k + 2 y^i D\beta \ . 
\end{equation}
$\beta$ is the coordinate on the Hopf fiber; we introduced covariant derivatives $Dy^i = dy^i + \epsilon^{ijk} y^j \omega^k$ and $D\beta = d\beta + A - \frac{1}{2} y^k \omega^k$.
$A$ is the Hopf connection satisfying: $d A =  -\frac{1}{2} {\rm vol}_{S^2}$.
Using the decomposition given in \eqref{mudec} we are finally able to rewrite the metric in Hopf coordinates as:
\begin{equation} \label{hopfmet}
ds^2_{3}=  D\beta^2 + \frac{1}{4} Dy^i Dy^i \ .
\end{equation}
We see that both the $S^1$ parametrized by $\beta$ and the $S^2$ parametrized by the $y^i$ are non-trivially fibered over $\Sigma_3$. For later convenience we also write the decomposition of the following two- and three-forms:
\begin{equation} \label{mumu}
\frac{1}{2} \epsilon^{ijk} \mu^{jk} = y^i \omega_1+ 2 Dy^i D\beta \ , 
\qquad \frac{1}{6} \epsilon^{ijk} \mu^{ijk} = 2 D\beta\ \omega_1 \ ,
\end{equation}
where the wedge products are implicit. As defined in (\ref{wi}), the two-form $\omega_1=\frac{1}{2} \epsilon^{ijk} y^i Dy^{jk}$ is the covariantized volume form of the $S^2$.

Finally notice that, when we reduce to ten dimensions along the Hopf fiber, the expression we got for $D\beta$ determines the one-form gauge field to be: $C_1=A- \frac{1}{2} y^k \omega^k$. The resulting RR two-form flux $F_2 = d C_1$ is then: $F_2 = \frac{1}{2} (\omega_5 - \omega_1)$, which is precisely the expression given in \eqref{fluxes30}.

\section{From ${\rm SU}(3)$ to \stt}
In this section we show how to decompose an SU(3) structure on the internal space $M_7$ to an \stt\ structure on $M_6$, where $M_7$ has the topology of an $S^1$ fibration over $M_6$ with the circle parametrized by the Hopf coordinate $\beta$.
This mapping is needed in order to give a complete proof that our 10d solution described in section \ref{natural} coincides in the massless limit with the reduction of the 11d solution described in \cite[Sec.~9.4]{gauntlett-macconamnha-mateos-waldram}. 

An SU(3) structure on $M_7$ is described by a real two-form $J$ and a complex three-form $\Omega$, that are given explicitly in reference \cite[Eq.(9.64)--(9.66)]{gauntlett-macconamnha-mateos-waldram}. In our notation\footnote{The eleven-dimensional metric corresponding to \eqref{433metric} can be written in ten dimensional language as
\begin{equation} 
ds^2_{11}= e^{-\frac{2}{3}\phi} \left( e^{2A}ds^2_{\rm AdS_4} +g^2 ds^2_{\Sigma_3}  + h^2 d\alpha^2 + f^2 Dy^i Dy^i \right) + e^{\frac{4}{3}\phi} D\beta^2 \ ,
\end{equation} 
where the explicit expressions for the functions entering the metric can be computed comparing this formula with \eqref{ads7s4n1}.}
these forms can be rewritten as:
\begin{align} 
J &= e^{-2 \phi /3} f g \mu^i e^i \ , \nonumber\\
\Re \Omega &= \frac{1}{6} f^2 \epsilon^{ijk} \mu^{ijk} - \frac{1}{2} g^2 \epsilon^{ijk} \mu^i e^{jk} \ , \\
\Im \Omega &= \frac{1}{2}f g \epsilon^{ijk} e^i \mu^{jk} - e^{-\phi} g^3 {\rm vol}_{\Sigma_3}\ . \nonumber
\end{align}
$f$ and $g$ here are the functions entering the 10d metric \eqref{433metric}, and the reduction from eleven to ten dimensions is performed as usual: $ds^2_{11}=e^{-2\phi/3}ds^2_{10}+ e^{4\phi/3}\left( d \beta + C_{1}\right)^2$.

We explained in appendix \ref{s3s2} how to decompose the forms $\mu^i$ defined on $S^3$ in terms of forms living on $S^2$ and on the Hopf fiber parametrized by $\beta$. In particular, we can use the relations \eqref{mudec}, \eqref{mumu} and decompose the SU(3) structure in terms of the twisted forms defined in section \ref{twistedforms} as:
\begin{equation}
\Omega=e^{-\phi} \left( 2 f D\beta + i g\ y^i e^i \right) \left( f^2 \omega_1 - g^2 \omega_5  + i f g \omega_2\right)\ , \qquad J = e^{-2\phi / 3} f g \left( 2 D\beta\ y^i e^i + \omega_4 \right)  \ . 
\end{equation}
Comparing this formula with \eqref{zjw}, it is clear that the SU(3) structure living on $M_7$ has been rewritten in terms of the SU(2) structure on $M_6$ as:
\begin{equation}
\Omega=e^{-\phi} \left( 2 f D\beta + i g\ y^i e^i \right) \left( \Im \omega -  ij \right)\ , \qquad J = e^{-2\phi / 3}  \left( 2 f g D\beta\ y^i e^i - \Re \omega \right)  \ . 
\end{equation}

This formula also allows to reduce the eleven-dimensional four-form flux $G_4$ given in \cite[Eq.(7.5)]{gauntlett-macconamnha-mateos-waldram}. As usual, $G_4 = F_4 + H \wedge D\beta$: the resulting RR four-form flux $F_4$ and NS three-form flux $H=dB$ coincide with the expressions we gave in \eqref{fluxes30}. 

\section{$J_\psi^{-1} \ \llcorner$ operator} \label{J-1}

In \cite[Sec.~5.2]{saracco-t}, the pure spinor equations \eqref{pse} were massaged for the particular case needed in this paper. All we need now is to compute the action of the $J_\psi^{-1}\ \llcorner$ operator on the two- and four-forms defined in section \ref{twistedforms}. $J_\psi^{-1}$ is a bi-vector defined as the inverse of the two-form $J_\psi$ entering the dielectric expression (\ref{dielectric}), which for our class of solutions can be expanded as: $ J_\psi = j_2 \omega_2 + j_3 \omega_3$, with coefficients $ j_2 = -\frac{f g}{\cos \psi}$ and $ j_3 = g $. 

It is natural to choose $f^i \equiv j_2 Dy^i - j_3 y^i dr\ $ as basis of one-forms on $M_3$ and the vielbein $e^i$  as basis on $\Sigma_3$, 
so that we can write $J_\psi$ as:
\begin{equation}
J_\psi = e^i \wedge f^i \ .
\end{equation}
Equivalently, the inverse operator can be expanded on the dual basis of vectors as:
\begin{equation}
J_\psi^{-1}\ \llcorner = F^i \llcorner E^i \llcorner \ ,
\end{equation}
where the basis of forms and dual vectors satisfy:
\begin{equation}
F^i \llcorner f^j = \delta^{ij} \ , \qquad F^i \llcorner e^j = 0 \ , \qquad E^i \llcorner f^j = 0 \ , \qquad E^i \llcorner e^j = \delta^{ij} \ .
\end{equation}
We now compute the dual vectors to be:
\begin{equation}
F^i = \frac{1}{j_2} v^i - \frac{1}{j_3} y^i dr \ , \qquad E^i = E^i_{0} - \epsilon^{jkl} v^j y^k ( E^i_{0} \llcorner \omega^l ) \ .
\end{equation}
$E^i_0$ are the dual vectors to $e^i$ on the base space satisfying $E^i_{0} \llcorner e^j = \delta^{ij}$. The vectors $v^i$ are given by
\begin{equation}
	v^1= \cos \theta \cos \varphi\, \del_\theta - \frac{\sin \varphi}{\sin \theta} \del_\varphi
	\ ,\qquad
	v^2 = \cos \theta \sin \varphi\, \del_\theta + \frac{\cos \varphi}{\sin \theta} \del_\varphi
	\ ,\qquad
	v^3= - \sin \theta \del_\theta \ ;
\end{equation}
they satisfy $v^i \llcorner Dy^j = \delta^{ij} - y^i y^j $. (They also happen to be conformal Killing vectors on $S^2$: ${\cal L}_{v^i}g_{S^2}= -2 y^i g_{S^2}$.) 

It is now straightforward to compute the action of $J_\psi^{-1}$ on the set of twisted two-forms:
\begin{equation}
\begin{split}
	&J_\psi^{-1}\ \llcorner \omega_1  = 0 \ ,\qquad
	J_\psi^{-1}\ \llcorner \omega_2  = \frac{2}{j_2} \ ,\qquad
	J_\psi^{-1}\ \llcorner \omega_3  = \frac{1}{j_3} \ , \\
	&J_\psi^{-1}\ \llcorner \omega_4  = 0 \ ,\qquad 
	J_\psi^{-1}\ \llcorner \omega_5  = 0 \ .	
\end{split}
\end{equation}
We finally compute the action of $J_\psi^{-1}$ on some the four-forms, which are also needed in the pure spinor equations:
\begin{equation}
J_\psi^{-1}\ \llcorner \omega_{13}  = \frac{1}{j_3} \omega_1 \ ,\qquad
J_\psi^{-1}\ \llcorner \omega_{15}  = \frac{1}{j_2} \omega_2 \ ,\qquad
J_\psi^{-1}\ \llcorner \omega_{35}  = \frac{1}{j_3} \omega_5 \ .
\end{equation}

\bibliography{at}
\bibliographystyle{at}

\end{document}